\documentclass[aps,prd,showpacs,showkeys,preprint,superscriptaddress]{revtex4}
\usepackage[caption = false]{subfig}
\usepackage[final]{graphicx}
\usepackage[usenames, dvipsnames]{color}
\usepackage{hyperref}
\usepackage{bm}
\usepackage{amsmath}
\usepackage{txfonts}
\usepackage{epstopdf}
\begin{document}
\title{Imprint of temporal envelope of ultrashort laser pulses on longitudinal momentum spectrum of $e^+e^-$ pairs}
\author{Chitradip Banerjee}
\email{banerjee.chitradip@gmail.com,   Present address: Department of Physics, Ariel University, Ariel 40700, Israel}
\author{Manoranjan P. Singh}
\affiliation{Theory and Simulations Lab, HRDS, Raja Ramanna Centre for Advanced Technology,  Indore-452013, India}
\affiliation{Homi Bhabha National Institute, Training School Complex, Anushakti Nagar, Mumbai 400094, India}

\begin{abstract}
Effect of temporal pulse shape of intense pulses on the momentum distribution of $e^+e^-$ pairs is studied using the quantum kinetic equation. Two closely resembling temporal envelopes namely, Gaussian and Sauter, keeping all the other pulse parameters same, are considered to this end. Contrary to the common perception which can be gauged from the interchangeable use of these temporal profiles, the longitudinal momentum spectrum of the pairs created by the two pulses is found to differ significantly in all the temporal regimes. 
For the pulses having a few cycles of oscillations, the temporal profile of the pulse is revealed in the oscillatory interference pattern riding over the otherwise smooth longitudinal momentum spectrum at asymptotic times. The onset of the oscillation due to the quantum interference of reflection amplitudes from the scattering potential due to the pulses having a temporal structure of multiple barriers takes place for a few-cycle oscillations for the Gaussian pulse than that for the Sauter pulse. Furthermore, the oscillation amplitude for the same number of oscillations within the pulse duration is larger for the Gaussian pulse. The presence of the carrier-envelope phase and the frequency chirping is found to magnify these differences. In the absence of any appreciable interference effect for the pulses having less than five oscillations, the longitudinal momentum spectrum has a higher peak value for the Sauter pulse at asymptotic times. On the hand, before the transient stage of evolution, the peak of the spectrum shows the opposite trend.
\end{abstract}

\maketitle

\section{introduction}
Particle-antiparticle pair production from the vacuum in the presence of an external gauge field or space-time curvature is one of the intriguing phenomena in the realm of quantum field theory \cite{SchutzholdUnruhHawking}. In the presence of the spatiotemporally homogeneous strong electric field, the generation of particle-antiparticle pairs is known as Schwinger mechanism \cite{PhysRev.82.664}. Another remarkable phenomenon of the black hole evaporation which includes the emission of all kinds of particles because of the very strong gravitational field of the black hole is known as Hawking radiation \cite{SWHawking1975}. The underlying physics behind these phenomena is that in the presence of the very strong background fields, the vacuum is not stable because of the quantum fluctuation and virtual particle-antiparticle pairs are separated by the energy of the background field or space-time causality to become a real one. According to Minkowski vacuum, there is no real particles for all observers/detectors at rest or moving at a constant velocity (via Lorentz invariance). However, yet another interesting quantum effect arises in the presence of constant acceleration. Here, an accelerating detector measures a thermal bath with temperature determined by the acceleration. The temperature is known as the Unruh temperature and the effect is known as the Unruh effect \cite{KimHawkingUnruh}.

 In principle, the generation of particle-antiparticle pair can be described either perturbatively in the process of highly energetic photons interacting with the heavy nucleus or non-perturbatively for the decay of vacuum in the presence of strong electric field \cite{RevModPhys.84.1177DiPiazza,BenKingReview}.
Despite being fundamentally important
the latter one lacks experimental verification because of its requirement of the peak field strength closer to $E_S = 1.32\times 10^{18}\textnormal{V/m}$ \cite{Sauter} (see the details of laser field parameters in \cite{RINGWALD2001107} ). Of late, the rapid development towards the generation of an ultraintense laser electromagnetic (EM) field, has inspired worldwide efforts to set up facilities such as the European Extreme Light Infrastructure for Nuclear Physics (ELI-NP)\cite{Heinzl2009EPJD,Dunne2009EPJD}, X-ray free electron laser (XFEL) at DESY, Hamburg using self amplified spontaneous emission (SASE) principle \cite{PhysRevLettDESYXFEL} for its experimental verification among other QED phenomena \cite{NaturePublishingGroupDunne,PhysRevLett.79.1626SLAC,RevModPhys.84.1177DiPiazza,Yanovsky:08}. In fact, nonlinear Compton scattering were observed in the collision of a $46.6$ GeV electron beam with terawatt laser pulses of $1053$ nm and $527$ nm with intensity $I = 10^{22}~\textnormal{W/cm}^2$ \cite{CbulaPhysRevLett.76.3116,DbrukePhysRevLett.79.1626} at SLAC.
 \\  
While the efforts are on for its experimental verification, the Schwinger mechanism was theoretically analyzed for various configurations of EM fields of the light sources currently available or expected to be available in the near future \cite{PhysRevLett.79.1626SLAC}. The theoretical description of Schwinger used electric field constant in time and uniform in space. The formalism was extended for time and space varying fields due to and ultraintense laser pulses \cite{PhysRevD.2.1191Itzykson,Hebenstreit2014189,Nikishov1970346,DynamiSchwingerBlaschke,Dunne2009EPJD,PhysRevD.80Chervyakov2009,PhysRevDHKleinert2008,PhysRevLettBulanov}. However, as the pulse duration is reduced further in the range of a few hundreds of attosecond when it is no more much larger than the characteristic Compton time, use of Schwinger formula to describe the pair production rate is questionable. Furthermore, in such cases, the transient and non-equilibrium dynamics of the produced particles can only be described in the framework of quantum kinetic equation(QKE)~\cite{PhysRevDSPKim2002,PhysRevDSPKim2007,PhysRevDSPKim2008,PhysRevDSPKim2009,AbdukerimCarrier,DynamiSchwingerBlaschke,Hebenstreit2014189,PhysRevD.82Dumlu,PhysRevD.73SPKim2006,PhysRevLettDumluStokes,BaiXIE2017225,BENEDETTI2013206,Filatov2008,PhysRevLett.67.2427Kluger,PhysRevD1994JRau,Smolyansky2012TimeReversalSymmetry,Smolyansky2017FieldPhaseTrans,BlaschkeCPP}. This methodology has strong relevance in the context of the momentum distribution of the created pairs in semiclassical approximation where the asymptotic reflection coefficient gives the average particle numbers in a particular mode \cite{PhysRevDDumlu2011}. The longitudinal momentum spectrum (LMS) of created particles has also been studied in nonpertubative multiphoton regime where the Keldysh adiabaticity parameter $\xi = m\omega_0/{eE} \sim 1$ for a given field strength $E$ and frequency $\omega_0$ \cite{PhysRevAMocken}.  The semiclassical formulation was used in studying the time-domain multiple-slit interference effect from vacuum in \cite{PhysRevLettRamseyFringes}. The kinetic theory was used to study the rich dynamical behaviour of the pair creation process for the time-dependent but spatially homogeneous field configuration \cite{BaiXIE2017225,AbdukerimCarrier,HEBENSTREIT2016PhysLettB}. In particular, it was shown that a quasi-particle mode evolves through three distinct temporal stages, namely the quasi electron-positron plasma (QEEP) stage, the transient stage and finally the residual electron-positron plasma (REPP) stage \cite{Smolyansky2012TimeReversalSymmetry,Smolyansky2017FieldPhaseTrans,BlaschkeCPP}.

The temporal characteristics of the pulses consist of a temporal profile with a given pulse duration $\tau$, number of oscillations $\omega \tau$ with $\omega$ being the carrier frequency, carrier-envelope phase (CEP) i.e., the phase difference between the high-frequency carrier wave and the temporal pulse envelope function and frequency chirp parameter(s). Gaussian and Sauter are the two most commonly used (quite often interchangeably) temporal profiles.  A simple Sauter pulse without any oscillation (also known as smooth Sauter pulse) offers analytical solutions for the momentum distribution \cite{PhysRevLett.67.2427Kluger} and the dynamics of produced pairs \cite{DynamiSchwingerBlaschke}. However, a Sauter pulse with a few-cycle oscillations is no longer analytically tractable. On the other hand, for a oscillating electric field Gaussian pulse it is possible to express the vector potential in an analytically closed form in terms of the error function. These analytical conveniences have led researchers to use Sauter and Gaussian temporal profiles for the kinetic studies of the pairs created by the smooth and oscillating electric field pulses, respectively \cite{PhysRevD.82Dumlu}, sometimes even in the same report \cite{Smolyansky2017FieldPhaseTrans}. This is possibly due to the perception that both the pulses should give very similar results because of their close resemblance. This, to the best of our knowledge, has not been verified so far. This is one of the motivations of this study.  While the evolution of individual modes was studied in Ref.~\cite{Smolyansky2017FieldPhaseTrans,BlaschkeCPP}, the evolution of the longitudinal momentum distribution as a whole has not been reported so far. This is the second motivation for our study. 
In this paper we, therefore, use QKE to present a detailed comparative study of the evolution of longitudinal momentum distribution of the pairs created by these two pulses (Sauter and Gaussian) for given pulse duration, number of few-cycle oscillations, CEP , and frequency chirp. 

We find that the LMS of the pairs for the Sauter and Gaussian pulses differ significantly at all the temporal stages of the evolution. However, for the qualitative description of the difference, only two temporal regions seem to be relevant - the first one is the region from the QEEP stage to the transient stage (referred to as the transient region hereafter) and the second one is the region well beyond the transient stage (also referred as the asymptotic region).

This paper is structured as follows: In Sec.~\ref{Theory} we discuss briefly the relevance of the aforesaid pulses to the counterpropagating configuration of intense ultrashort pulses. We also outline the basic formulation of QKE in the context of particle production from the time-dependent but spatially uniform electric fields. We present our numerical results for the oscillating electric field with Sauter and Gaussian pulses with different values of $\omega\tau$ parameter in Sec.~\ref{Results}. The effect of varying the CEP and the linear frequency chirp on the LMS is also studied in this section. The results are qualitatively explained by invoking the equivalence between the pair creation by EM field and the over-the-barrier scattering problem and also quantitatively by analysing the structure of turning points in the complex $t$-plane in stationary phase approximations. Although our study pertains to the tunneling regime of pair production we also briefly present the effect of temporal profile on the LMS in the multi-photon regime of pair production towards the end of the section. We conclude in Sec.~\ref{conclusion}. Details of the calculation based on the turning point structure showing the essential difference in the LMS of the two temporal pulse forms are relegated to Appendix.~\ref{Appendix}.

\section{theory}\label{Theory}

\subsection{Electric field model}\label{EM_field_model}
In order to study the effect of temporal envelope on the LMS of created pairs we consider a spatially homogeneous electric field 
\begin{equation}\label{Eq:Electric_field}
E(t) = E_0g(t)\sin(\omega t+\phi),\quad H(t) = 0,
\end{equation}
where $g(t)$ is the temporal envelope function to describe the electric field of a finite duration and $\phi$ is the CEP \cite{RevModPhys_CEP,RevModPhys_CEP1_nonlinear,Dipiazza_CEP} which plays an important role in laser matter interaction for the ultrashort pulses. Now we discuss for the pulse profile with time-independent frequency i.e., $\omega$ is constant. 
If we take $g(t) = \exp(-t^2/{2\tau^2})$, we get the spatially uniform but temporally oscillating electric field  with Gaussian profile
$
E(t) = E_0\exp(-t^2/{2\tau^2})\sin(\omega t+\phi),
$
where $\tau$ is the total pulse duration and its product with the frequency $\omega$, i.e. $\omega \tau$ gives the number of oscillations within the pulse envelope.
On the other hand, $g(t) = \cosh^{-2}(t/\tau)$ represents the oscillating electric field  with Sauter pulse profile
$
E(t) = E_0\cosh^{-2}(t/\tau)\sin(\omega t+\phi).
$

It should be noted here that a possible method of achieving such an electric field is to use two counterpropagating laser beams when both of them are either linearly e-polarized or circularly polarized in right-left combination \cite{ChitradipCEP,ChitradipPhasePhysRevA}. 
\begin{figure}[t]
   \begin{center}
   \includegraphics[width = 3.0in]{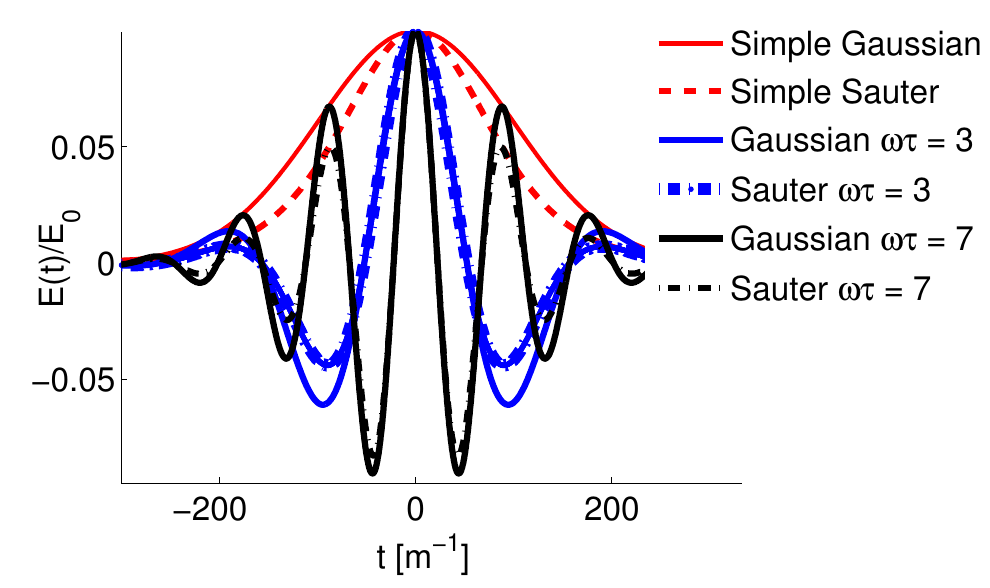}
   \caption{Plot of oscillating electric field with Sauter (dashed line) and  Gaussian (simple line) pulses  for the number of oscillations $\omega\tau = 3$ (blue) to $\omega\tau = 7$ (black). The smooth Sauter and Gaussian fields (red) are given as a reference. The field parameters are $E_0 = 0.1$, $\tau = 100$, $\phi = \pi/2$ and all the units are taken in the  electron mass unit.}
   \label{EtSaugaus}
   \end{center}
   \end{figure}
Fig.~\ref{EtSaugaus} shows the shape of the time dependent electric field for Sauter and Gaussian pulses with different values of $\omega\tau$ parameter. The value of the $\omega\tau$ parameter is taken as $3$ and $7$. The shape of the smooth Sauter ($E(t) = E_0\cosh^{-2}(t/\tau)$) and Gaussian ($E(t) = E_0\exp(-t^2/{2\tau^2})$) fields are shown as a reference. We use the Eq.~\ref{Eq:Electric_field} for the oscillating electric fields to calculate the LMS of created particles.

\subsection{Frequency chirping for the finite electric pulse with harmonic oscillations}

 Now we discuss the EM field with frequency chirping. Suppose we have plane EM wave propagation along $z$-direction with wave-vector $k\hat{z}$, for which we can write the electric and magnetic fields as
 $$
 {\bf{E}}({\bf{r}},t)={\bf{E}}_0({\bf{r}}_{\bot})\exp(-i\psi)
 $$
 and 
 $$
 {\bf{B}}({\bf{r}},t)={\bf{B}}_0({\bf{r}}_{\bot})\exp(-i\psi).
 $$
 Here ${\bf{E}}_0({\bf{r}}_{\bot})$ and ${\bf{B}}_0({\bf{r}}_{\bot})$ are constant and $\psi=\omega_0t-kz=\omega_0(t-z/c)$ where $k=\omega_0/c$. We further have $\psi=\omega_0(t-z)$, where $c=1$(in natural unit) is taken. Now to get the finite pulsed electric and magnetic fields we follow the transformation in the plane wave factor as
 $$
 \exp(-i\psi)\rightarrow if'(\psi),
 $$
  where $f(\psi)=g(\psi/{\omega_0\tau})\exp(-i\psi)$ and prime denotes the derivative with respect to the phase $\psi$. This kind of substitution satisfies the Maxwell's equations up to the orders of $1/{\omega_0\tau}$ where $\omega_0$ is the central frequency of the pulse with pulse duration $\tau$. The envelope $g(\psi/{\omega_0 t})$ is assumed to be $1$ at the centre of the pulse, $g(0)=1$, and exponentially decreasing for $|\psi|\gg \omega_0\tau$.
 Now if we consider the frequency chirping  (here we concentrate up to linear frequency chirping), which we have by the following substitution into the phase factor as 
 $$
 \psi \rightarrow \psi'=\frac{\psi}{1-b\psi}
 $$
 for $|b\psi|< 1$. Up to linear order of $\psi$ we have
 $$
 \psi'=\psi+b\psi^2.
 $$
 Therefore we have
 $$
 f(\psi')=g\left(\frac{\psi+b\psi^2}{\omega_t\tau}\right)\exp\left(-i(\psi+b\psi^2)\right)
 $$
 Here we are interested in the $z=0$ plane i.e., in a plane passing through the focal centre for which we have $\psi'=\omega_tt-kz=\omega_t t=\omega_0 t+b\omega_0^2t^2$ i.e., $\omega_t=\omega_0+b\omega_0^2t$. So if we do the above substitution we can rewrite $f(\psi')$ as
 $$
 f(\psi')=g\left(\frac{\omega_tt}{\omega_t\tau}\right)\exp\left(-i(\omega_0+b\omega_0^2t)t\right)=g\left(\frac{t}{\tau}\right)\exp\left(-i(\omega_0+b\omega_0^2t)t\right).
$$
In Sec.\ref{Result_Chirp} we define $\beta=b\omega_0^2$ is the linear frequency chirp parameter.
\subsection{Quantum Kinetic Equation}
The theory of pair production due to the Schwinger mechanism is non-perturbative. One has to, therefore, strive for finding an exact solution of the Dirac equation and quantum kinetic theory is one such description, particularly well suited for the case of pair production being affected by the ultrashort laser pulses \cite{PhysRevD1999Schmidt,PhysRevD1994JRau,PhysRevLett.67.2427Kluger}. 
In the presence of the time-dependent but spatially uniform electric field, the vacuum state evolves to a state where mixing between the positive and negative energy states takes place. Enormous analytical simplification results by the  diagonalization of the underlying  Hamiltonian through the Bogoliubov transformation in the quasi-particle basis 
\begin{equation}
H(t) = \sum\limits_{s,\textbf{p}}\omega(\textbf{p},t)\Big(B^{\dagger}_{\textbf{p},s}(t)B_{\textbf{p},s}(t)- D_{-\textbf{p},s}(t)D^{\dagger}_{-\textbf{p},s}(t)\Big),
\end{equation}
 with $\omega^2({\bf{p}},t) = \epsilon^2_{\perp}+P_3^2(t)$ is the total energy squared of the quasiparticle. Here $\epsilon^2_{\perp} = m^2+{\bf{p}}_{\perp}^2$ is the transverse energy squared and $P_3(t) = p_3-eA(t)$ is the $z$-component of time dependent kinematic momentum which is responsible for the particle acceleration $dP_3(t)/{dt} = eE(t)$; $e$ is the electronic charge. $B_{\textbf{p},s}(t)$, $B_{\textbf{p},s}^{\dagger}(t)$ and $D_{\textbf{p},s}(t)$, $D_{\textbf{p},s}^{\dagger}(t)$ are the slowly varying part of the time dependent creation and annihilation operators for quasi-particles and anti-(quasi)particles respectively, see for example Ref.~\cite{Schmidth1998IJMPE}. These operators satisfy the following Heisenberg like equation of motion: 
\begin{equation}\label{Heisenberg like Equation}
\begin{split}
\frac{dB_{\textbf{p},s}(t)}{dt} = -\frac{eE(t)\epsilon_{\perp}}{2\omega^2(\textbf{p},t)}D^{\dagger}_{-\textbf{p},s}(t)+i[H(t),B_{\textbf{p},s}(t)],\\
\frac{dD_{\textbf{p},s}(t)}{dt} = \frac{eE(t)\epsilon_{\perp}}{2\omega^2(\textbf{p},t)}B^{\dagger}_{-\textbf{p},s}(t)+i[H(t),D_{\textbf{p},s}(t)].
\end{split}
\end{equation} 
The quantum kinetic theory deals with the evolution of the quasiparticle distribution function which is  defined as $f_s(\textbf{p},t) = <0_{in}|B^{\dagger}_{\textbf{p} s}(t)B_{\textbf{p} s}(t)|0_{in}>$ \cite{Schmidth1998IJMPE,DynamiSchwingerBlaschke}. A straightforward use of the Hisenberg-like equation of motion in Eq.~\ref{Heisenberg like Equation} results in the following dynamical equation for the distribution function  $f_s(\textbf{p},t)$ 
\begin{equation}
\frac{df_s(\textbf{p},t)}{dt} = -\frac{eE(t)\epsilon_{\perp}}{2\omega^2(\textbf{p},t)}Re\{\phi_s(\textbf{p},t)\},
\end{equation}
where the function $\phi_s(\textbf{p},t) = <0_{in}|D_{-\textbf{p} s}(t)B_{\textbf{p} s}(t)|0_{in}>$ gives a measure of  particle-antiparticle correlation is  the complex order parameter describing the evolution of the initial vacuum state, see Ref.~\cite{PhysRevD.100Chitradip}. It satisfies the equation
\begin{equation}\label{Eq:OrderParameter}
\frac{d\phi_s(\textbf{p},t)}{dt} = \frac{eE(t)\epsilon_{\perp}}{\omega^2(\textbf{p},t)}[2f_s(\textbf{p},t) -1]-2i\omega(\textbf{p},t)\phi_s(\textbf{p},t).
\end{equation}
Since the distribution function and the particle-antiparticle correlation function do not depend on the spin $s$, we can drop the spin index $s$ in $f_s(\textbf{p},t)$ and $\phi_s(\textbf{p},t)$. By introducing two auxiliary functions $u(\textbf{p},t)= -\rm Re[\phi(\textbf{p},t)]$ which gives the measure of the vacuum polarization,  and $v(\textbf{p},t)=\rm Im[\phi(\textbf{p},t)]$ which provides measure of the counter reaction of the vacuum in response to the strong external gauge field, QKE can be written in the form of three coupled first order differential equation as
\begin{eqnarray}\label{Set_ODE_KE}
  \frac{df(\textbf{p},t)}{dt} &=& \frac{eE(t)\epsilon_{\perp}}{2\omega^2(\textbf{p},t)}u(\textbf{p},t),\nonumber\\
  \frac{du(\textbf{p},t)}{dt} &=&\frac{eE(t)\epsilon_{\perp}}{\omega^2(\textbf{p},t)}[1-2f(\textbf{p},t)]-2\omega(\textbf{p},t)v(\textbf{p},t),\nonumber\\
  \frac{dv(\textbf{p},t)}{dt} &=& 2\omega(\textbf{p},t)u(\textbf{p},t).
  \end{eqnarray}
This set of coupled differential equations for $f(\textbf{p},t), u(\textbf{p},t),~\textnormal{and}~v(\textbf{p},t)$ has the first integral of motion $(1-2f(\textbf{p},t))^2+ u(\textbf{p},t)^2 + v(\textbf{p},t)^2 = 1$  \cite{PhysRevD.90.125033Huet}. It may be noted here that the above QKE are obtained by ignoring the effect of created pairs on the external electric field and the collisional effects of the pairs. These equations can be cast into a single  first-order integro-differential equation:
\begin{equation}\label{QKE_integro_diff_Eqn}
  \frac{df(\textbf{p},t)}{dt} = \frac{eE(t)\epsilon_{\perp}^2}{2\omega^2(\textbf{p},t)}\int\limits_{-\infty}^{t}dt^{\prime} \frac{eE(t^{\prime})}{\omega^2(\textbf{p},t^{\prime})}[1-2f(\textbf{p},t^{\prime})]\cos[2\Theta(\textbf{p};t,t^{\prime})]. 
   \end{equation}  
 with $\Theta({\bf{p}}; t_1, t_2) = \int\limits_{t_1}^{t_2}dt^{\prime}\omega({\bf{p}}, t^{\prime})$ being the dynamical phases accumulated between initial and final states in the presence of the time dependent gauge field. This form of QKE (Eq.~\ref{QKE_integro_diff_Eqn}) has non-local time structure i.e. $df(\textbf{p},t)/{dt}$ depends on all the time history it passes through by the term $(1-2f(\textbf{p},t^{\prime}))$ which carries the quantum statistical information due to the Pauli exclusion principle and the highly oscillating phase kernel $\cos[2\Theta(\textbf{p};t,t^{\prime})]$ and hence not amenable to  numerical methods. 
We, therefore, solve Eq.~\ref{Set_ODE_KE} numerically for time dependent smooth and oscillating electric field with ultrashort Sauter and Gaussian pulse profiles as discussed in Sec.~\ref{EM_field_model} with the initial conditions $f(\textbf{p},t_i) = u(\textbf{p},t_i) = v(\textbf{p},t_i) = 0$ as $t_i \rightarrow -\infty$ and present our results in the Sec.~\ref{Results}.

\section{Results}\label{Results}
In this section, we present the LMS of the created particle based on the numerical solution of Eq.~\ref{Set_ODE_KE} for time-dependent Sauter and Gaussian pulses, thereby showing its sensitivity to the temporal envelope of the pulse.

\subsection{LMS of created particles in the transient and asymptotic regions}

The virtual electron-positron pairs in the initial vacuum state interact with the electric field. However, initially, the pairs are in the off-shell mass configuration. As time progresses the pairs acquire energy from the electric field and move towards the on-shell mass configuration. In this process the distribution function exhibits three distinct dynamical stages \cite{otto2016,Smolyansky:2016gmp,Nousch:2016yxy,Smolyansky:2014dna,Panferov:2016bln}. In the first stage which is known as QEPP stage, the distribution function shows oscillations of the electric field. In the second stage, the distribution function shows rapid oscillations. In this stage which is known as the transient stage, the dynamics is governed by the electric field, the vector potential and the accumulated phase through the highly oscillating kernel which can be easily seen from the kinetic equation as given by Eq.~\ref{QKE_integro_diff_Eqn}. This is why the oscillations in the transient stage are so rapid. The transient stage is signified by particle reaching the on-shell configuration from initial off-shell mass configuration \cite{Filatov2008}. As mentioned in the previous subsection the electron-positron pairs are correlated which is quantified  by the complex order parameter $\phi(\textbf{p},t)$. The distribution function $f(\textbf{p},t)$ undergoing rapid oscillations in the transient stage was attributed to a sudden increase in phase of complex order parameter $\phi(\textbf{p},t)$ \cite{PhysRevD.100Chitradip} which results in the dephasing in the correlation between quasiparticle pairs. After this dephasing, the electron-positron pairs turn into ``independent particles''. This stage is known as REPP stage \cite{Smolyansky2012TimeReversalSymmetry,Smolyansky2017FieldPhaseTrans}.  

The evolution for smooth Gaussian pulse has not been reported so far while that for the oscillating electric field with Sauter pulses was reported recently \cite{PhysRevD.100Chitradip}. However, there are no results available for the pulse parameters of both the pulses we wish to consider in this paper.

\begin{figure}[t]
\begin{center}
{\includegraphics[width = 3.2in]{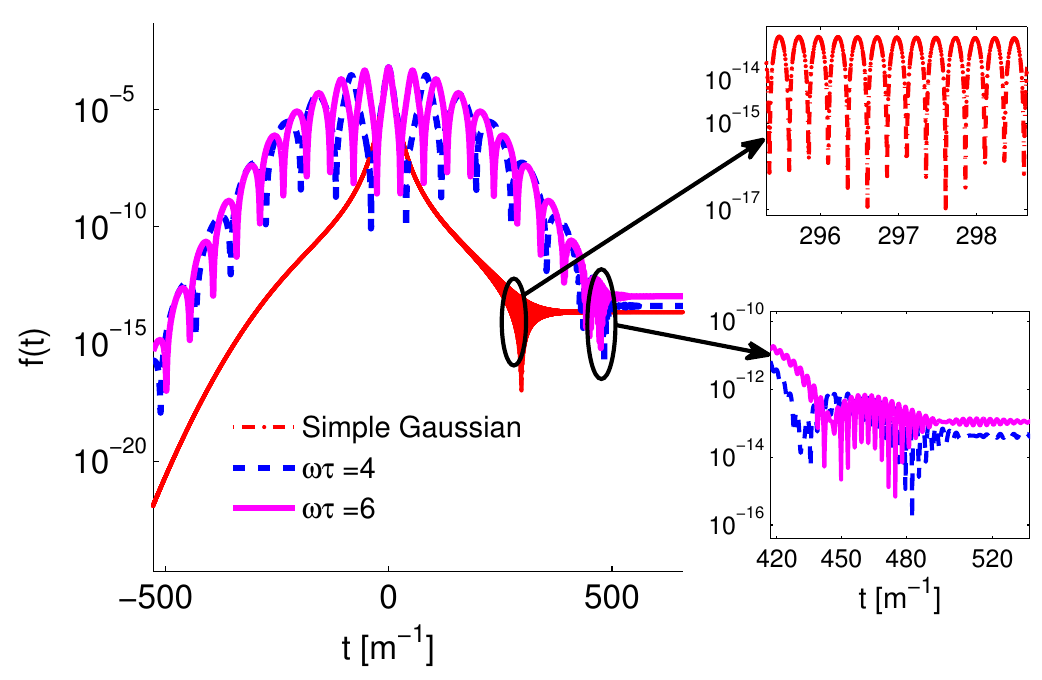}
\includegraphics[width = 3.2in]{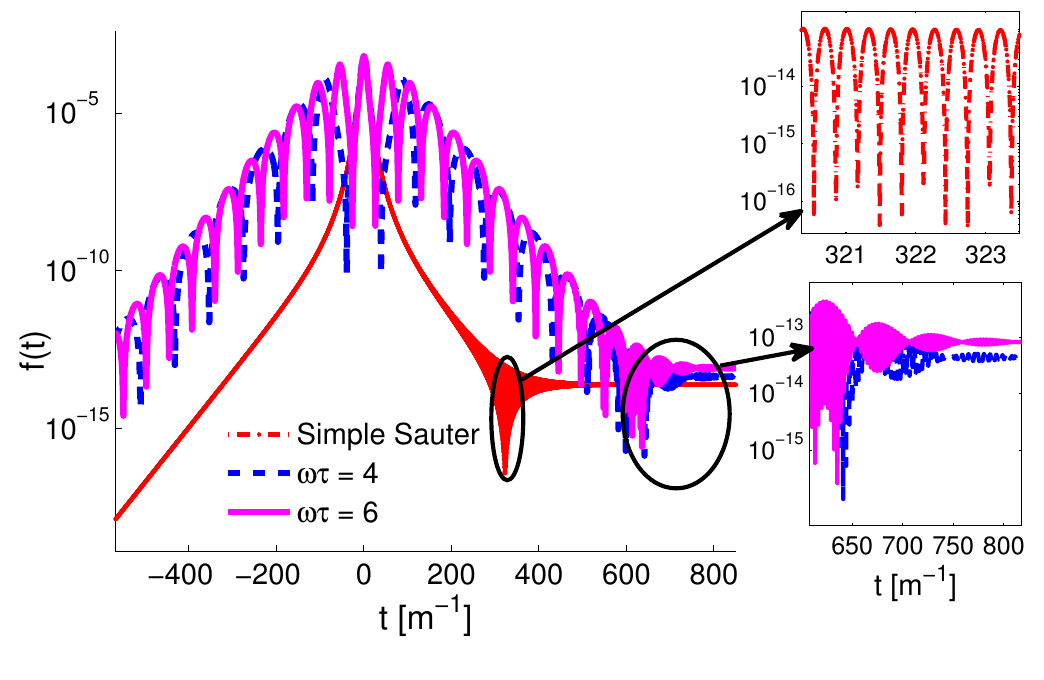}}
\caption{Evolution of quasi-particle distribution function for smooth and oscillating electric field  ($\omega\tau = 4, 6$) with Gaussian (on the left) and Sauter (on the right) pulses for longitudinal momentum $p_3 = 0$ and transverse momentum $\textbf{p}_{\perp}=0$. All the units are taken in electrons mass unit. The field parameters are $E_0 = 0.1$, $\tau = 100$, and the CEP $\phi = \pi/2$.}
\label{FtGaussp0OmTau046}
\end{center}
\end{figure}
We, therefore, show the complete evolution of the quasi-particle distribution function for the smooth and oscillating electric field  ($\omega\tau = 4, 6$) with Gaussian (in the left panel) and Sauter (right panel) pulses in  Fig.~\ref{FtGaussp0OmTau046} for the longitudinal momentum value $p_3 = 0$. The insets of Fig.~\ref{FtGaussp0OmTau046} show the evolution of the quasi-particle distribution function in the transient stage of evolution characterized by rapid oscillations. We note here that the transient stage occurs at earlier times for the Gaussian pulse than the Sauter pulse.

\begin{figure}[t]
\begin{center}
\includegraphics[width = 4.6in]{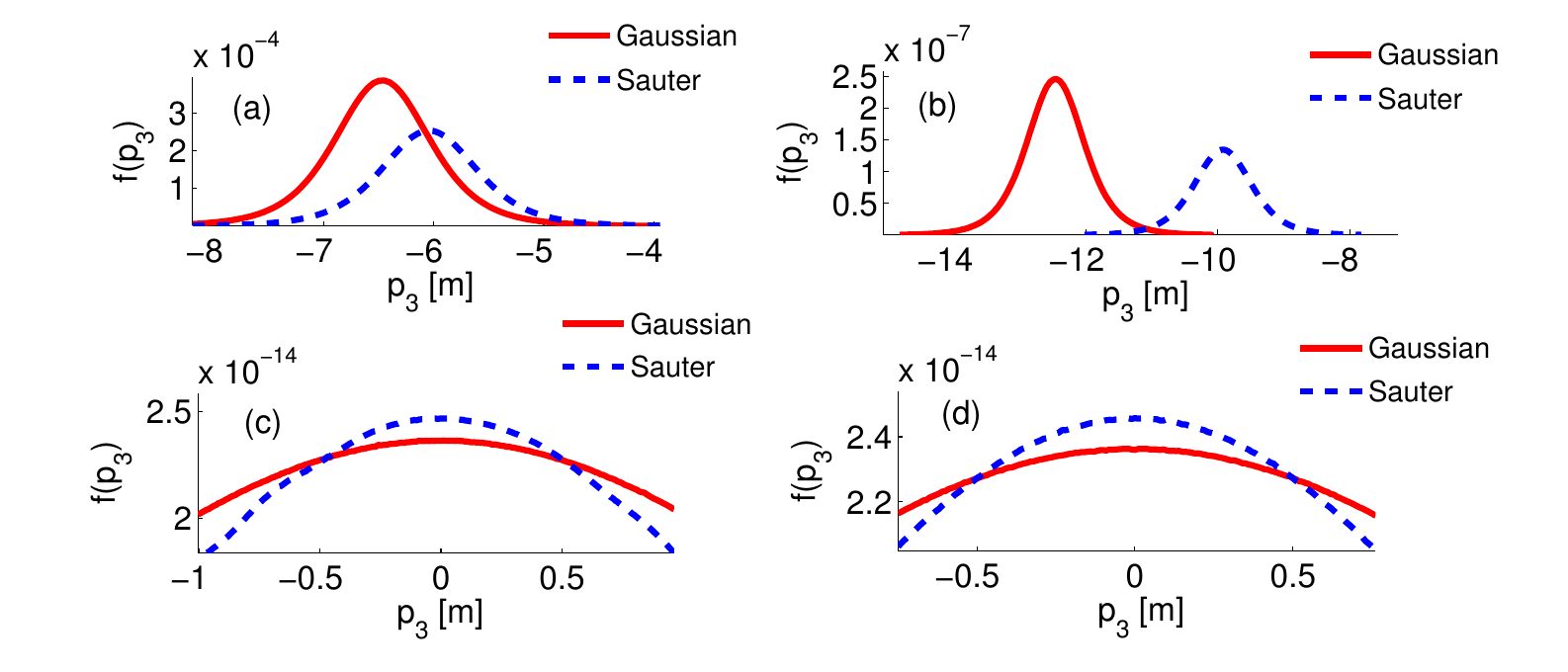} 
\caption{ LMS of created particles in the presence of time dependent smooth Sauter and Gaussian pulses at different times. (a) $t = 70$. (b) $t = 280$. (c) $t = 560 $. (d) $t=1050$. The value of transverse momentum is taken to be zero and all the units are taken in electrons mass unit. The field parameters are $E_0 = 0.1$, $\tau = 100$, and the CEP $\phi = \pi/2$.}
\label{SimpleSauGausFP3T560}
\end{center}
\end{figure}

Use of QKE formalism allows us to study the quasi-particle LMS at any instant of time and hence the evolution thereof. As mentioned before we, however, restrict ourself to two distinct temporal regimes, namely the transient regime (consisting of QEPP and transient stages) and the asymptotic region (well beyond the transient stage in REPP stage). Fig.~\ref{SimpleSauGausFP3T560} shows momentum spectra for smooth Sauter and Gaussian pules at $t = 70,~ 280,~ 560,$ and $1050$. The spectra, which have a smooth unimodal structure, change rapidly in the transient region. The location of the central peak and the peak height of the LMS are quite different for both the pulses. In particular, the spectrum for the Gaussian pulse has a larger peak height than that for the Sauter pulse in this region, as seen in  Fig.~\ref{SimpleSauGausFP3T560} (a) and (b).  However in the asymptotic region the momentum spectra for both the pulses become centrally symmetric about $p_3 = 0$, see  Fig.~\ref{SimpleSauGausFP3T560} (c) and (d). Contrary to the trend in the transient region, the peak height is larger for the Sauter pulse in the asymptotic region. For the smooth Sauter pulse with electric field $E(t) = E_0 \cosh^{-2}( t/\tau)$ and the corresponding vector potential $A(t) = -E_0\tau\tanh(t/\tau)$, QKE in Eq.~\ref{QKE_integro_diff_Eqn} was shown to have exact solution in the asymptotic region \cite{PhysRevDKluger1998}. It was shown \cite{PhysRevDDumlu2011} that in the stationary phase approximation, the asymptotic time spectrum is governed by the structure of the turning points $t_p$ in the complex time plane  which are defined by the relation $\omega( \textbf{p},t_p) = 0$. In this case, $t_p = \tau \tanh^{-1}((\pm im-p_3)/{eE_0\tau})+in\pi\tau$ where $n$ is an integer. The turning points appear as complex conjugate pairs. For $p_3 = 0$, all the turning points are located on imaginary axis. The dominant contribution to the LMS comes from the $n = 0$ turning points. Thus, the pair creation mechanism is governed by the single pair of the turning points and the asymptotic particle spectrum has a smooth unimodal profile. 

\begin{figure}[t]
\begin{center}
{\includegraphics[width = 2.0in]{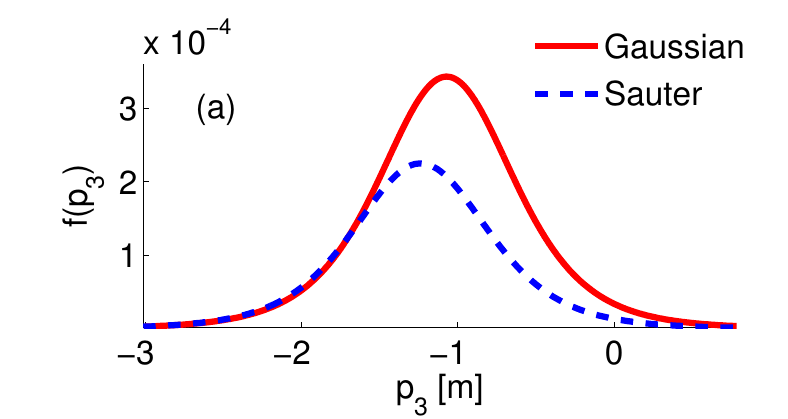}
\includegraphics[width = 2.0in]{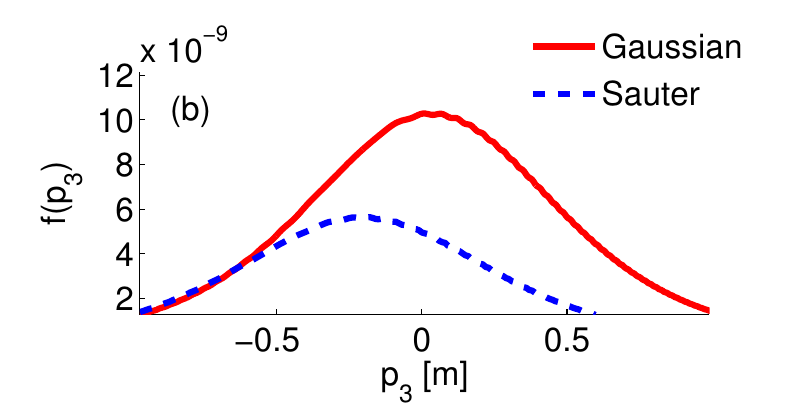}
\includegraphics[width = 2.0in]{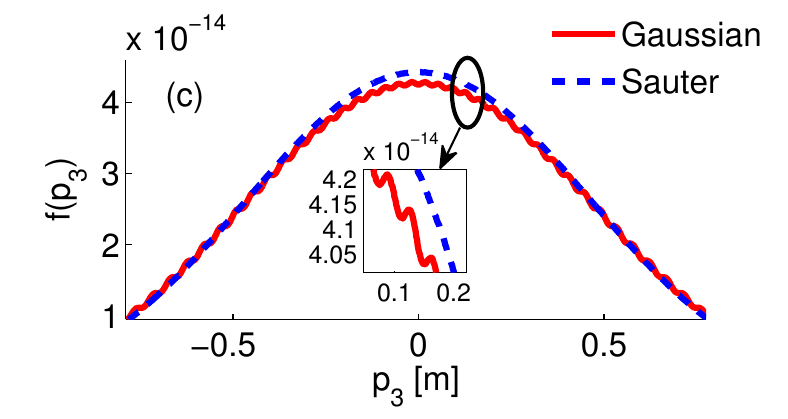}\\
\includegraphics[width = 2.0in]{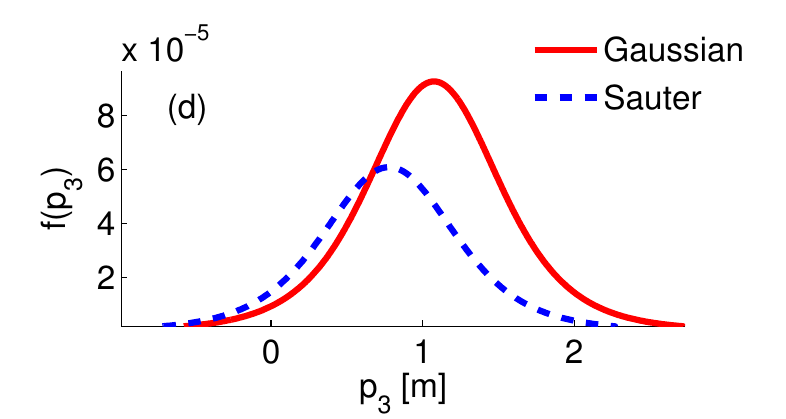}
\includegraphics[width = 2.0in]{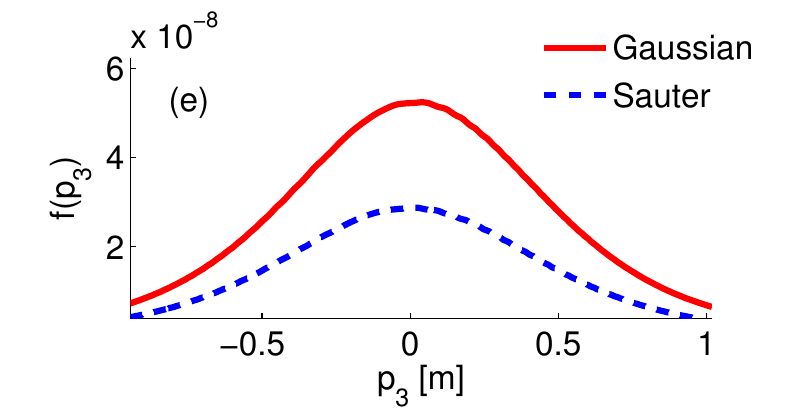}
\includegraphics[width = 2.0in]{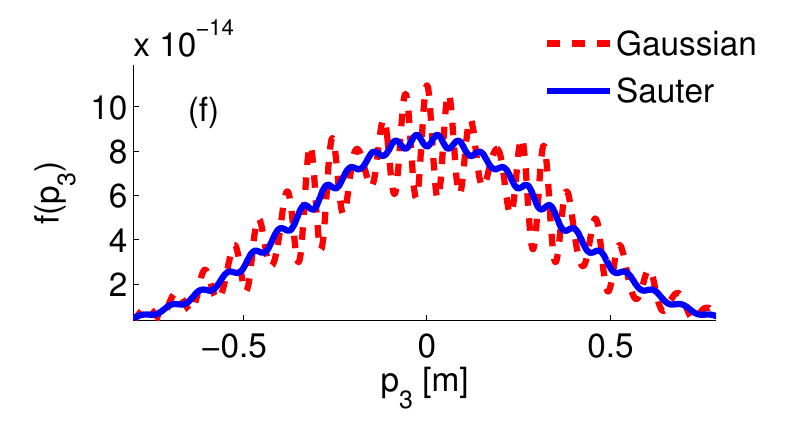}}
\caption{LMS of created particles in the presence of time dependent oscillating electric field with Sauter and Gaussian pulse profiles at different times. (a)$t = 70$, $\omega \tau = 4$. (b)$t = 280$, $\omega \tau = 4$. (c)$t = 1050$, $\omega \tau = 4$. (d)$t = 70$, $\omega \tau = 6$. (e)$t = 280$, $\omega \tau = 6$. (f)$t = 1050$, $\omega \tau = 6$. The value of transverse momentum is taken to be zero and all the units are taken in electrons mass unit. The field parameters are $E_0 = 0.1$, $\tau = 100$, and the CEP $\phi = \pi/2$.}
   \label{GausSauomTau46T280}
   \end{center}
   \end{figure}

\begin{figure}[t]
   \begin{center}
   {\includegraphics[width = 2.0in]{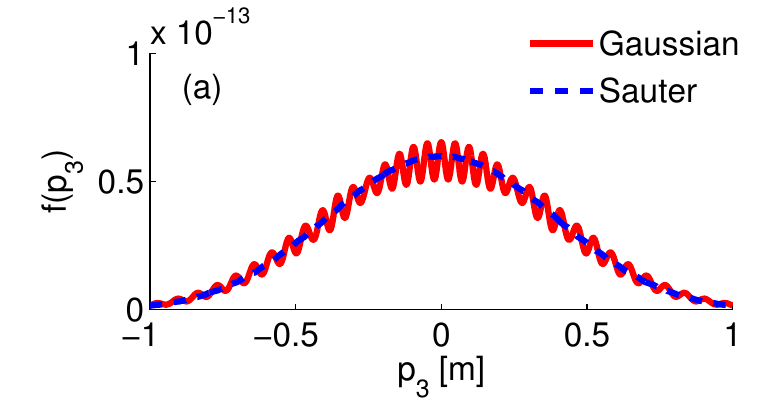}
    \includegraphics[width = 2.0in]{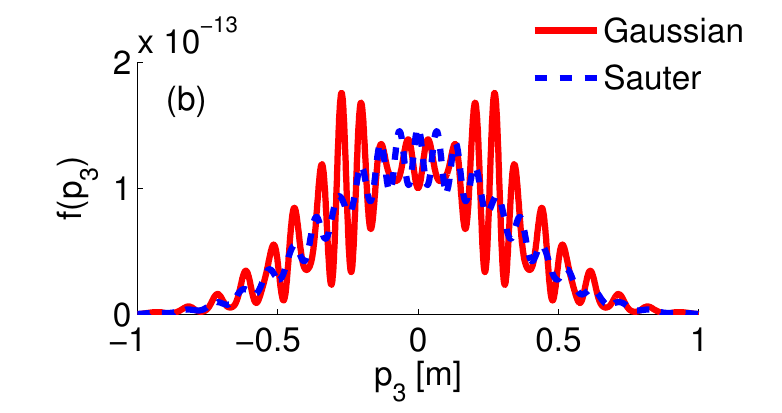}
    \includegraphics[width = 2.0in]{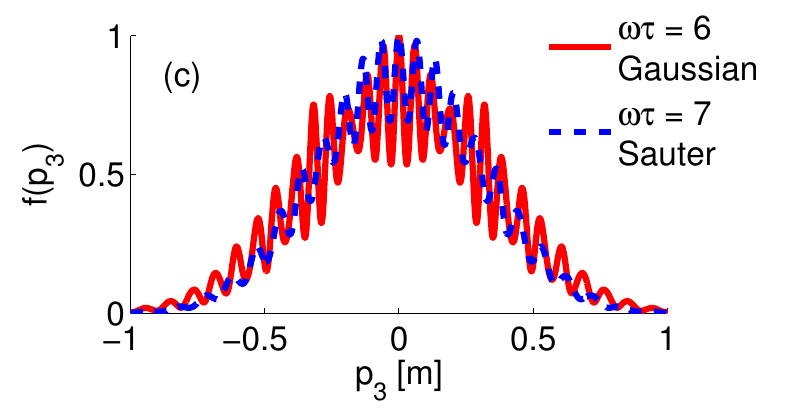}}
   \caption{Asymptotic longitudinal momentum distribution function (calculated at $t = 10.5\tau$) of the created particles in the presence of time dependent oscillating electric field with Sauter and Gaussian pulse profiles for different values of $\omega \tau$ parameter. (a) $\omega\tau = 5$, (b) $\omega\tau = 7$ for both the pulses. (c) The distribution functions scaled with respect to the respective peak values for the Gaussian pulse with $\omega\tau = 6$ and for the Sauter pulse with $\omega\tau = 7$. The value of transverse momentum is taken to be zero and all the units are taken in electrons mass unit. The field parameters are $E_0 = 0.1$, $\tau = 100$, and the CEP $\phi = \pi/2$.}
   \label{SauterMomOmTau}
   \end{center}
   \end{figure}

We now consider the oscillating electric field with Sauter and Gaussian pulse profiles for $\omega\tau = 4$ and $6$ and present in Fig.~\ref{GausSauomTau46T280} the momentum spectra in the transient region ($t = 70,~280 $) and in the asymptotic region ($t = 1050$. The momentum spectra in the transient region, much like the trend discussed above,  significantly differ for the two pulses for the same number of oscillations, with the peak height for the Sauter pulse being consistently lower than that for the Gaussian pulse. In the asymtotic region, oscillations over the otherwise smooth unimodal LMS for the oscillating electric field with Gaussian pulse were reported in \cite{PhysRevLettHebenstreitMomentum, PhysRevDDumlu2011}. These results are reproduced here for the ready reference here while comparing the results obtained for the corresponding Sauter pulses.  We find that the oscillation  becomes noticeable for $\omega \tau \geq 4$ for the Gaussian pulse and  for $\omega \tau \geq 6$ for the Sauter pulse, see the lower panel of the Fig.~\ref{GausSauomTau46T280} and also Fig.~\ref{SauterMomOmTau}. 
 
For the same number of oscillations, the amplitude of oscillations is larger for the Gaussian pulse than that for the Sauter pulse, as seen in Fig.~\ref{SauterMomOmTau}. In fact, the spectrum for the $\omega\tau = 6$ Gaussian pulse is similar to that for the Sauter pulse with $\omega\tau = 7$ (Fig.~\ref{SauterMomOmTau}) as far as the oscillations in both cases are concerned (note the scaling factors). As long as the oscillations are not prominent ($\omega\tau \leq 5$) the spectrum has higher peak value for the Sauter pulse than that for the Gaussian pulse.  With the increase in the oscillation amplitude in the LMS for Gaussian pulses the trend reverses for ($\omega\tau \geq 6$). 
 
\subsection{Scattering potential structure for the Gaussian and Sauter pulses}\label{Scattering potential}
\begin{figure}[t]
          \begin{center}
          {\includegraphics[width = 2.0in]{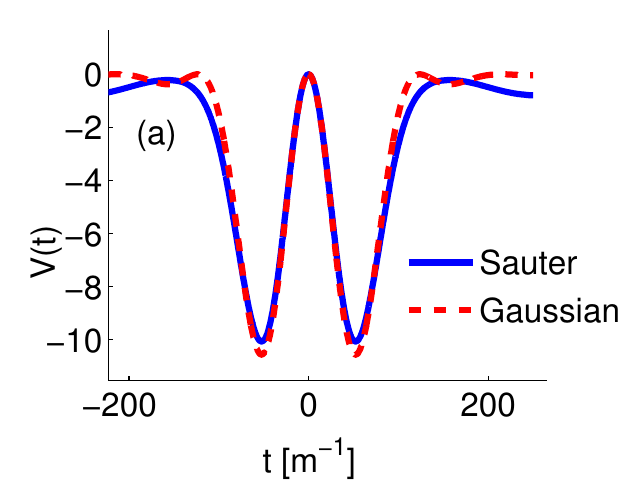}~
            \includegraphics[width = 2.0in]{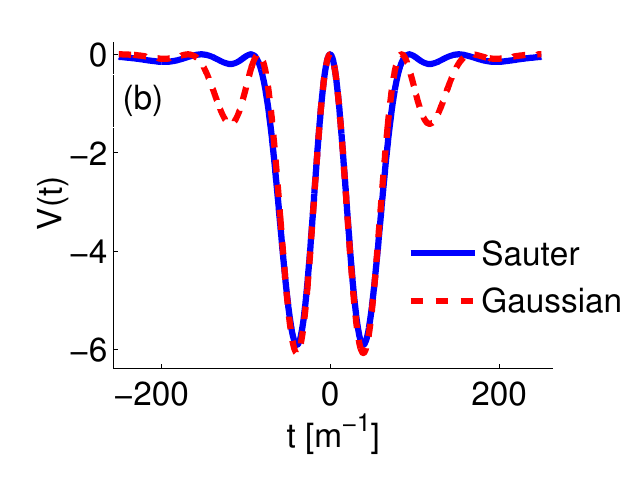}~
            \includegraphics[width = 2.0in]{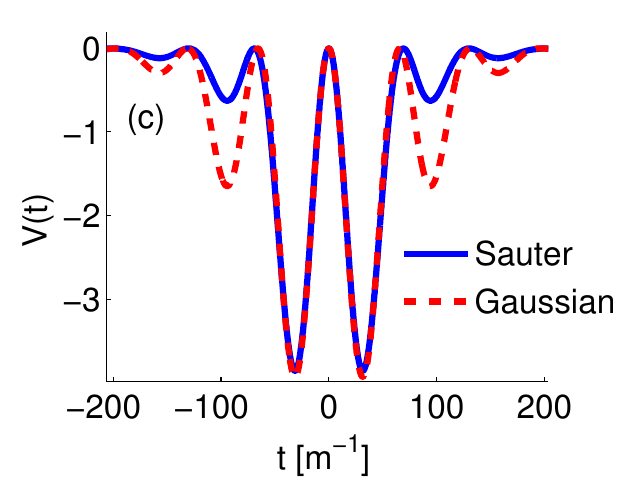}\\
            \includegraphics[width = 2.0in]{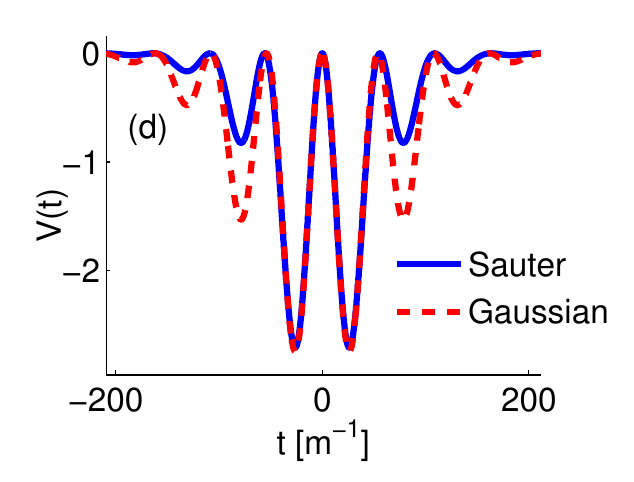}~
            \includegraphics[width = 2.0in]{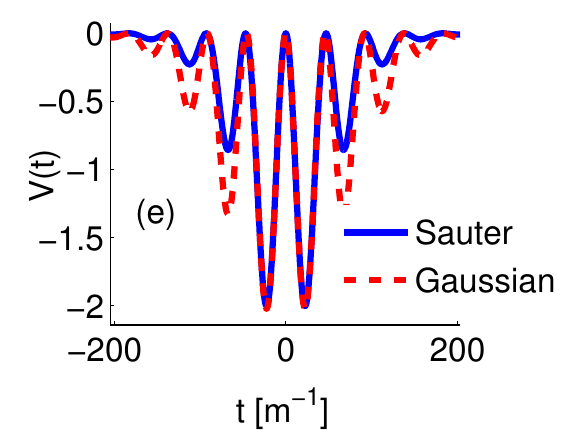}~
            \includegraphics[width = 2.0in]{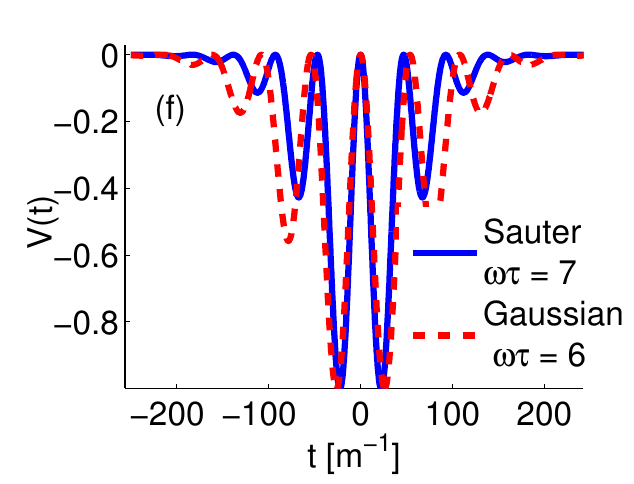}}
          \caption{Over-the-barrier scattering potential $V(t) = -(p_3 - eA(t))^2$ for time dependent oscillating electric field with Gaussian and Sauter  pulse profiles. From top left to bottom middle (a-e) the value of the $\omega \tau = 3, 4, 5, 6, 7$.  The scaled scattering potentials due to the Sauter pulse with $\omega\tau = 7$ and the Gaussian pulse with $\omega\tau = 6$ are shown on the bottom right (f). Longitudinal momentum $p_3 = 0$. All units are expressed in electrons mass unit. The field parameters are $E_0 = 0.1$, $\tau = 100$, and the CEP $\phi = \pi/2$.}
            \label{potentialOmTauSauGaus}
            \end{center}
\end{figure}

The physical explanation of the onset of oscillation over the otherwise smooth LMS induced due to the oscillating electric field  within the Gaussian pulse was provided in \cite{PhysRevD.82Dumlu} by mapping the problem of pair creation by the spatially uniform time-dependent pulses to the well studied over the barrier scattering problem of quantum mechanics.  
It was shown that the pairs creation is related to the reflection of the initial $t\rightarrow -\infty$ quasi particle mode with longitudinal momentum $p_3$ (transverse momentum $\textbf{p}_{\perp}=0$ without any loss of generality) at asymptotic times $t\rightarrow \infty$ due to the time dependent scattering  potential, $V(t) = -(p_3-eA(t))^2$. We use the same physical picture to explain the suppression of oscillations in the LMS of the pairs created by the Sauter pulse compared to that by the corresponding Gaussian pulse. In Fig.~\ref{potentialOmTauSauGaus}, we show the scattering potential $V(t) = -(p_3-eA(t))^2$ for both smooth and oscillating electric field with Sauter and Gaussian pulse profiles with $\omega\tau = 3, 4, 5, 6, 7$ for the longitudinal momentum $p_3 = 0$. The potential is symmetric about $t = 0$.  It is smooth, having a single bump (or barrier) for the smooth and the oscillating electric field pulses having a small number of oscillations, $\omega\tau \leq 3$ (out of all these $V(t)$ for only the oscillating electric field pulses with $\omega \tau =3$ is shown here). As the value of $\omega\tau$ is increased the structure of the potential gets more bumpy causing multiple reflections of the incident wave. It is the interference of the multiple reflected waves which results in the oscillations in the LMS at asymptotic times. For $\omega\tau = 4$, the scattering potential, as shown in Fig.~\ref{potentialOmTauSauGaus}(a) has three bumps -- the larger one in the centre and two smaller ones on either side of the central one towards the tail region of the pulse. For the Sauter pulse, on the other hand, the scattering potential has one bump in the centre, with the side bumps being hardly visible.  Hence, the LMS for the Sauter pulse with $\omega\tau = 4$ in Fig.~\ref{GausSauomTau46T280} does not show the interference effects. On the other hand, the spectrum for the corresponding Gaussian pulse in Fig.~\ref{GausSauomTau46T280} does show mild interference effect in the form of very small amplitude oscillations over the smooth unimodal profile. 

With the increase in the number of oscillations, the existing side bumps become more prominent and additional side bumps appear in the scattering potential. In all the cases, however, the side bumps for the Sauter pulse is less prominent than those for the corresponding Gaussian pulse. The relative strength of the prominent side bumps with respect to the central one is nearly $1/3$ for the Sauter pulse and $1/2$ for the Gaussian pulse with  $\omega\tau = 6$. For $\omega\tau = 7$, it is about $1/2$ for the Sauter pulse and $2/3$ for the Gaussian pulse. This explains the relative suppression of oscillations in the LMS for the Sauter pulse. 
 
The side bumps for the Sauter pulse become visible for $\omega\tau = 5$ (Fig.~\ref{potentialOmTauSauGaus}(b)) but these are too small to cause any discernible interference effects in the LMS. The onset of oscillation in the LMS for the Sauter pulse takes place only at $\omega\tau = 6$. In Fig.~\ref{potentialOmTauSauGaus}, we compare the scattering potential due to the Sauter pulse with $\omega\tau = 7$ with that due to the Gaussian pulse with $\omega\tau = 6$. The similarity of the two potential structures explains the similarity in the LMS for the two pulses with a different number of oscillations.

It may be helpful to relate the difference in the scattering potential due to Sauter and Gaussian pulses to their respective electric field profiles. As seen in Fig.~\ref{EtSaugaus}, for higher values of $\omega \tau$, the oscillations located away from the centre of the temporal envelope are more intense for the Gaussian pulse making thereby the corresponding side bumps more prominent.

\subsection{Turning point structure for Sauter and Gaussian pulses}\label{Turning point}
\begin{figure}[t]
       \begin{center}
            \includegraphics[width =6.2in]{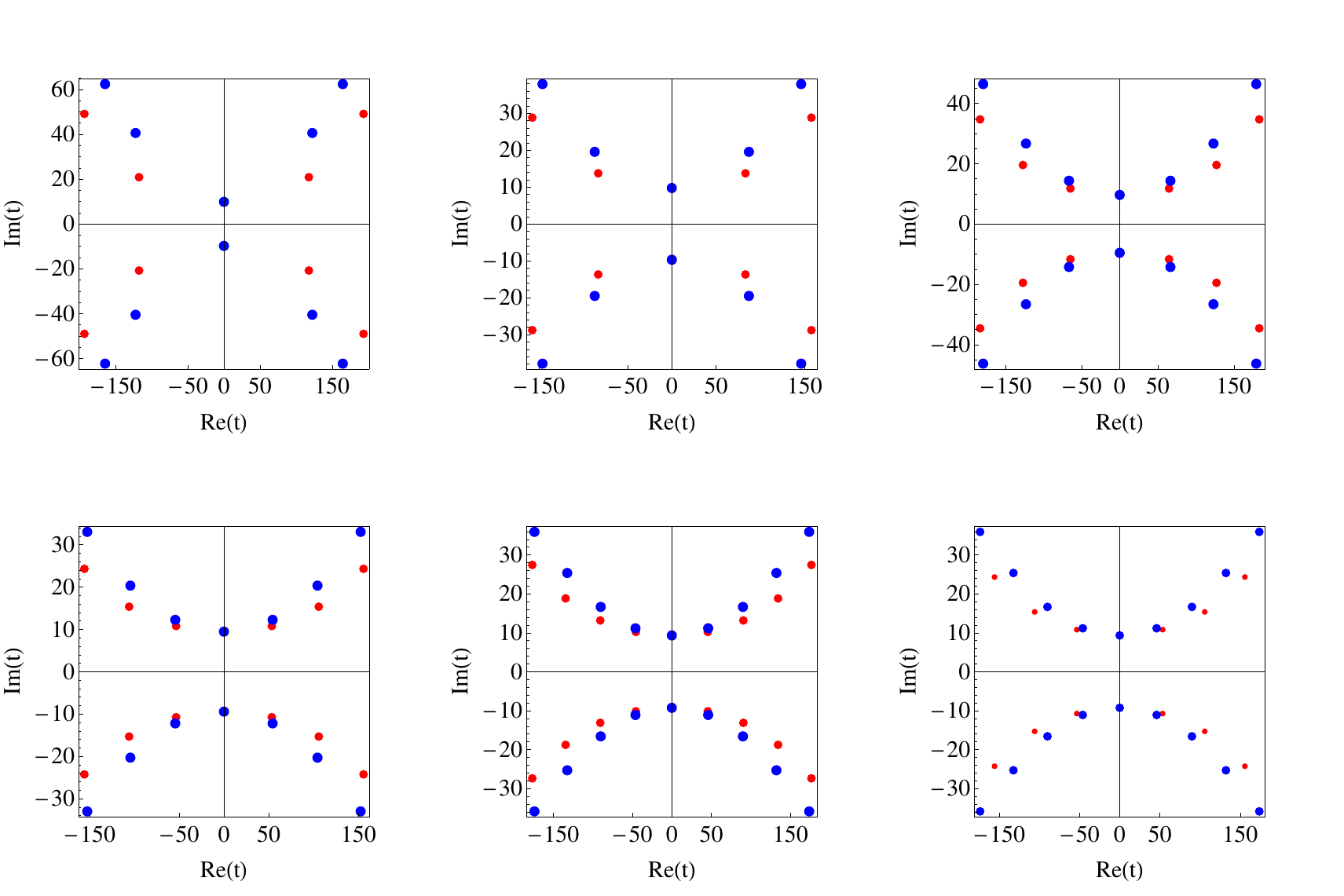} 
            \caption{Turning points $t_p$ in complex $t$-plane for the oscillating electric field with Sauter (blue dots) and Gaussian pulses (red dots) with same number of oscillations. $\omega \tau = $3, 4, 5, 6, and 7 from top left to bottom middle. The plot at the bottom right shows the turning points for Sauter pulse with $\omega \tau = 7$ and Gaussian pulse with $\omega \tau = 6$. Longitudinal momentum $p_3 = 0$. Transverse momentum $p_{\perp} = 0$. All units are expressed in electrons mass unit. The field parameters are $E_0 = 0.1$, $\tau = 100$, and the CEP $\phi = \pi/2$.}
            \label{GauSauterTurning1May2018}
            \end{center}
\end{figure}

Mapping the calculation of the reflection amplitude of the over barrier scattering problem to finding the poles in the complex t-plane has been gainfully employed to solve many problems in physics since a long time, starting with L. Landau \cite{LandauQuantum}, V L Pokrovsky \cite{Pokrovsky2010}, Brezin-Itzykson \cite{PhysRevD.2.1191Itzykson}. In \cite{PhysRevD.82Dumlu,PhysRevDDumlu2011}, this theoretical framework was used to calculate the LMS of the created pairs by the spatially uniform time-dependent pulses. In particular, it was shown that for a subcritical field reflection amplitude in the asymptotic time limit   $R_{\textbf{p}}(\infty)$, satisfying the condition $|R_{\textbf{p}}(\infty)|\ll 1$ can be expressed as a sum involving all the turning points:
\begin{equation}\label{Eq:R_asymptotic}
R_{\textbf{p}}(\infty) \approx \sum\limits_{t_p}(-1)^pe^{i\pi/2}e^{-2i\int\limits_{-\infty}^{t_p}dt^{\prime}\omega(\textbf{p}, t^{\prime})}.
\end{equation}
As mentioned earlier, turning points $t_p$ are defined in the complex $t$-plane by the relation  $\omega({\textbf{p}}, t_p) = 0$. They  appear in complex conjugate pairs as  the vector potential $A(t)$ considered here is real. It was argued in Ref. \cite{PhysRevLettDumluStokes} that the function  $\exp(-2i\int\limits_{-\infty}^{t_p}dt^{\prime}\omega(\textbf{p}, t^{\prime}))$ is  oscillatory along the real axis of the complex $t$-plane and  exponentially decaying along the imaginary axis. Thus, only the pairs of turning points located near the real axis contribute significantly to the reflection amplitude and the corresponding  terms in the expression of reflection amplitude represent the reflection due to the significant bumps/barriers of the scattering potential discussed in the previous subsection. If the reflection is governed by more than a single pair of turning points, the resulting LMS of pairs will show interference effects in form of oscillations. For definiteness, the asymptotic momentum distribution $f_{\textbf{p}}(\infty) = |R_{\textbf{p}}(\infty)|^2$ is given by \cite{PhysRevDDumlu2011}
\begin{equation}\label{eq:N_k(inf)}
f_{\textbf{p}}(\infty) \approx \sum\limits_{t_p}e^{-2K^{(p)}_{\textbf{p}}}+2\sum\limits_{t_{p^{\prime}} \neq t_p} (-1)^{(p-p^{\prime})}\cos\bigg(2\Theta^{(p,p^{\prime})}_\textbf{p}\bigg)~e^{-K^{(p)}_{\textbf{p}}-K^{(p^{\prime})}_{\textbf{p}}},
\end{equation}
where $K^{(p)}_{\textbf{p}} = \bigg|\int\limits_{t^*_p}^{t_p}dt~\omega(\textbf{p}, t)\bigg|$ and $\Theta^{(p,p^{\prime})}_\textbf{p} = \int\limits_{Re(t_p)}^{Re(t_{p^{\prime}})}dt ~\omega(\textbf{p}, t)$.
In  Eq.~\ref{eq:N_k(inf)}, the first term on the right hand side contains the contribution of reflections from all the pairs of turning points and the second term represents the inference of reflected waves from different pairs of turning points.  Because of the exponential suppression factor $\exp(-2K^{(p)}_{\textbf{p}})$ the dominant contribution to the asymptotic distribution function comes from those pairs of turning points which are closer to the real axis. Therefore,  a closer look into the structures of the turning points in Fig.~\ref{GauSauterTurning1May2018} for both pulses, will be able to shed light on the nature of the resulting LMS in the asymptotic region. 
As mentioned earlier, for $p_3=0$ (and $p_{\perp} = 0$, considered for convenience), all the turning points are located on the imaginary axis for the smooth Sauter pulse. The separation between successive turning points is $\pi\tau$ which is enormous. The pair creation, therefore, is dictated by the single pair of turning point which is closest to the real axis and is given by $t_p = \tau \tanh^{-1}((\pm im-p_3)/{eE_0\tau})=\pm 9.96687 i$. Turning points for the corresponding Gaussian pulse are very close to those for the Sauter pulse, with the relevant pair closest to the real axis being $\pm 9.98339 i$. Thus in both the cases, the asymptotic LMS has smooth unimodal profile. Since the turning point for the Sauter pulses is slightly closer to the real axis, the resulting LMS has somewhat larger peak value as shown is Fig.~\ref{SimpleSauGausFP3T560}. 

We present in Fig.~\ref{GauSauterTurning1May2018} the turning point structure for the oscillating electric field with pulse profiles with $\omega\tau =  3, 4, 5, 6$ and $7$ for the longitudinal momentum mode $p_3 = 0$. For the oscillating electric field pulses, besides the central turning point pair on the imaginary axis, there are other pairs of turning points symmetrically located on either side of the imaginary axis. Although these turning points are located much closer to the real axis compared to the $n \neq 0$ turning points for the smooth pulses, for the values of  $\omega\tau \leq 3$ they are still far-off to give any significant contribution to the reflection amplitude. As the value of $\omega\tau$ is increased, more pairs of turning points have comparable imaginary values of $t$, thereby giving rise to the possibility of the interference effects in the asymptotic reflection coefficient and hence the oscillatory pattern in the LMS of created particles. For the Gaussian pulse with $\omega\tau = 4$, there are two such pairs symmetrically located on either side of the central pair, at a distance of $83.1$ unit along the real axis (note that the side bumps of the scattering potential also appear close to these locations, see Fig.~\ref{potentialOmTauSauGaus}). The distance of these turning points from the real axis is $13.7068$ unit which is comparable to the distance of $9.7359$ unit of the central pair. For the Sauter pulse, the corresponding additional turning point pairs are located at a distance of $19.5236$ units which is more than twice the distance at which the central turning point pair is located. This explains the suppression of oscillations in the LMS for the Sauter pulse. It is only at $\omega\tau = 6$ that the non-central turning point pairs are located at a distance ($12.2396$ units) from the real axis which is comparable to that for the central turning point pair. These turning points cause oscillations in the LMS. Appendix ~\ref{Appendix} contains a detailed calculation of interference effect in the LMS which brings out that the onset of oscillations for the Gaussian pulse takes place for $\omega\tau = 4$ whereas for the Sauter pulse, oscillations start at $\omega\tau = 6$.

For the same number of oscillations within the pulse duration, the amplitude of oscillations in the LMS is larger for the Gaussian pulse as the turning points causing the interference lie closer to the real axis in this case than those for the Sauter pulse. The turning point structure for the Gaussian pulse with $\omega\tau = 6$ and the Sauter pulse with $\omega\tau = 7$ are quite similar as shown in Fig.~\ref{GauSauterTurning1May2018} (bottom right). This is consistent with the similarity of the scaled scattering potentials for the two pulses, as shown in Fig.~\ref{potentialOmTauSauGaus} (bottom right) and also explains strong resemblance between oscillations in the respective momentum spectra (Fig.~\ref{SauterMomOmTau}).

The central pair of turning points for the Sauter pulse is always slightly closer to the real axis than that for the Gaussian pulse. Therefore, unless the interference effect due to the other pairs of turning points becomes strong enough, the LMS for Sauter pulse has a higher peak value.

\subsection{LMS for Sauter and Gaussian pules with variation in CEP}
Thus far, the CEP $\phi$ is taken to be $\pi/2$ in order to match the maximum of the oscillating electric field to the maximum of the temporal envelope. In this case, both Sauter and Gaussian pulses are nearly identical in the central part. Therefore, the overall profile and the centre of the momentum spectra discussed so far are quite similar for both the pulses. The noticeable difference between the two pulses in the non-central region is reflected in the suppression of oscillations in the LMS for the Sauter pulse. For other values of $\phi$, the maximum of the electric field is pushed away from the central region towards one of the tail regions, making it asymmetrically distributed in the pulse envelope. It is, therefore, expected that variation in $\phi$ will make the difference in the momentum spectra for the Sauter and Gaussian pulses more pronounced. 
  
\begin{figure}[t]
\begin{center}
\includegraphics[width = 2.8in]{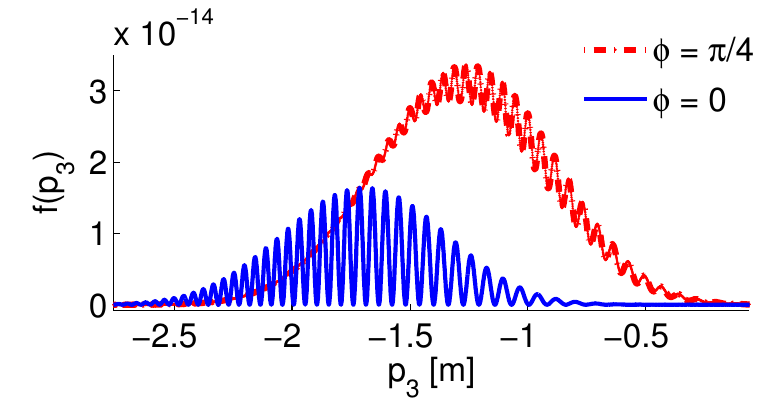} 
\caption{Asymptotic distribution function of the created particles in the presence of time dependent oscillating electric field with Sauter pulse profile as a function of longitudinal momentum of the particles with different values of CEP $(\phi)$. The transverse momentum is taken to be zero and all the units are taken in electrons mass unit. The field parameters are $E_0 = 0.1$, $\tau = 100$, and $\omega = 0.05$.}
\label{FPSauOmTau5CEP}
\end{center}
\end{figure}

The dashed line of Fig.~\ref{FPSauOmTau5CEP} shows the LMS of the created particles for the Sauter pulse with $\omega\tau = 5$ with the CEP $\phi = \pi/4$. The corresponding spectrum for Gaussian pulse was reported in Fig.~3 of Ref.~\cite{PhysRevLettHebenstreitMomentum}.  In the spectrum for the Sauter pulse the oscillations are drastically suppressed, the peak value is lower, the width is larger and the centre is at $p_3= -1.262$ which correspond kinetic momentum $P_3(\infty) = 130.816$ keV. Note that the spectrum for the Gaussian pulse has maximum kinetic momentum $P_3(\infty)=102$ keV. 

\begin{figure}[t]
\begin{center}
{\includegraphics[width = 2.4in]{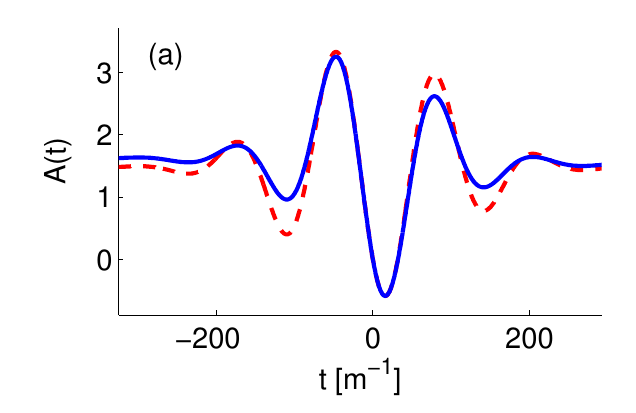}~
            \includegraphics[width = 2.4in]{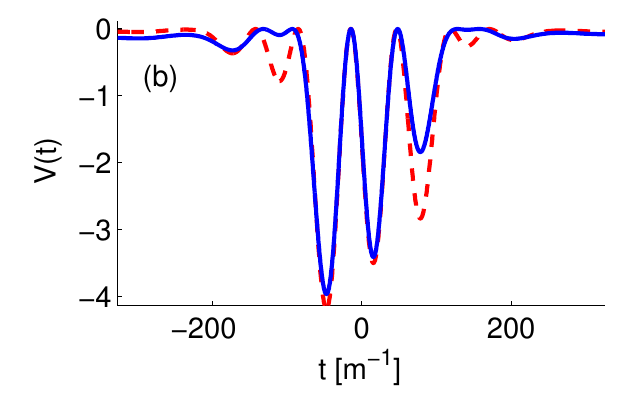}}
\caption{Plot of $A(t)$ and $V(t)$ for Gaussian (dashed line) and Sauter (solid line) with $\phi = \pi/4$. The $V(t) = -(p_3-eA(t))^2$ is plotted for the longitudinal momentum values $p_3 = -1.286$ for the Gaussian pulse and $p_3 = -1.262$ for the Sauter pulse. All the units are taken in electrons mass unit. The field parameters are $E_0 = 0.1$, $\tau = 100$, and $\omega = 0.05$.}
\label{VACEP}
\end{center}
\end{figure}

\begin{figure}[t]
          \begin{center}
           {\includegraphics[width = 2.8in]{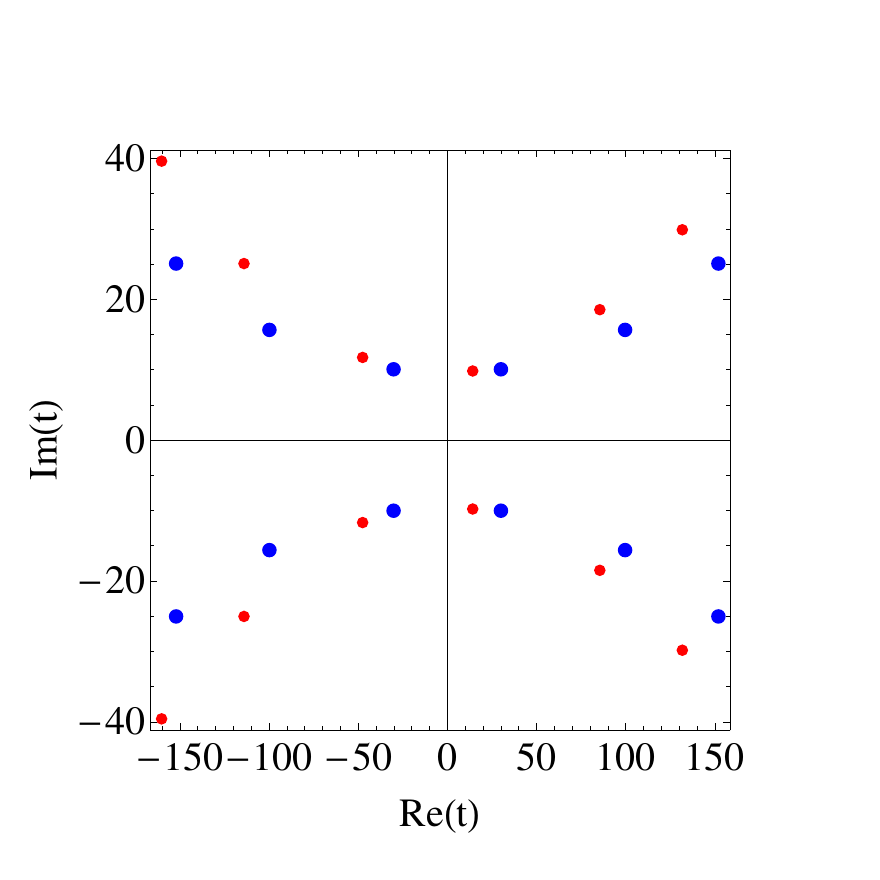}~
            \includegraphics[width = 2.8in]{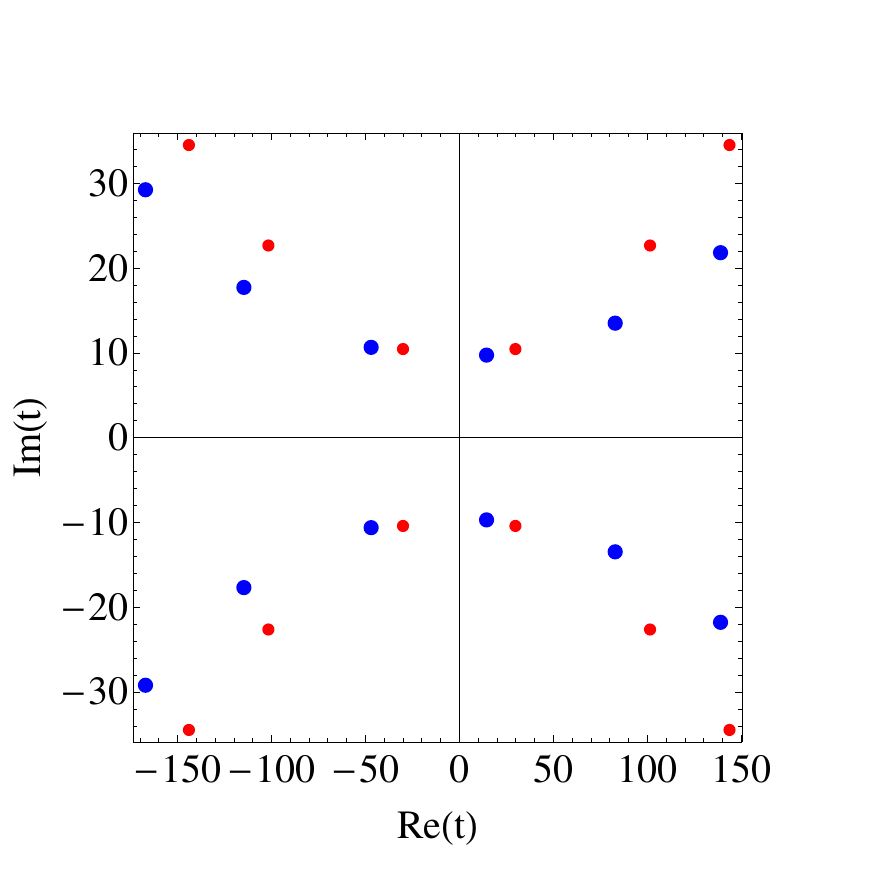}}
    \caption{Turning points $t_p$ in complex $t$-plane for the oscillating electric field with Sauter pulse (small dots) and Gaussian pulse profiles (large dots) with same number of oscillation ($\omega\tau = 5$) for CEP $\phi = \pi/4$ (left panel) and $\phi = 0$ (right panel). Longitudinal momentum $p_3 = -1.262$ and $p_3 = -1.286$ for Sauter and Gaussian pulses respectively with $\phi = \pi/4$ and $p_3 = -1.714$ and $p_3 = -1.830$ for the same with $\phi = 0$. All units are expressed in electrons mass unit. The field parameters are $E_0 = 0.1$, $\tau = 100$.}
             \label{TurningCEP4590}
            \end{center}
\end{figure}

To understand the origin of these differences we plot vector potential for $\phi = \pi/4$ for both pulses as shown in Fig.~\ref{VACEP}(a). The respective scattering potentials using the values of $p_3$ at which the spectra are centred are shown Fig.~\ref{VACEP}(b). These, as expected, are asymmetric. The dominant potential barrier located at $t=14.2$ is nearly the same for both the pulses. The additional barriers on either side of this barrier, located at $t=-48.2$ and $t=89$,  are significantly weaker for the Sauter pulse, suppressing thereby the interference effects in the reflection coefficient and hence the oscillations in the LMS of created pairs. The explanation can be substantiated by looking at the turning point structure for both the pulses with $\phi=\pi/4$ as shown in the left panel of Fig.~\ref{TurningCEP4590}. We take the longitudinal momentum $p_3 = -1.262$ for the Sauter pulse and $p_3 = -1.286$ for the Gaussian pulse. It shows that for the Gaussian pulse, the dominant turning points are located at $t_{p_1}=-29.31+i 9.466$ and $t_{p_2}=31.45+i 9.939$ whereas for the Sauter pulse these points are located at $t_{p_1}=-46.13+i 11.83$ and $t_{p_2}=15.56+i 9.466$. As for the Sauter pulse, the $\rm Im(t_{p_1})=11.83$ and $\rm Im(t_{p_2})=9.466$, the main contribution in the pair production  comes from the tunnelling from $t_{p_2}$ to $t^*_{p_2}$ and the oscillations which are caused by the interference between the neighbouring turning points are suppressed by the exponential factor. However, for the Gaussian pulse, the imaginary value of the turning points are close enough and we observe strong oscillations as shown in Ref.~\cite{PhysRevLettHebenstreitMomentum}. 

In Fig.~\ref{FPSauOmTau5CEP}, the LMS for the Sauter pulse with $\phi = 0$ should be compared with that reported for Gaussian pulse in Fig.~4 of Ref.~\cite{PhysRevLettHebenstreitMomentum}. All the aforesaid differences in the momentum spectra are once again seen here. The feature of the LMS can be explained by looking at the turning point structure for both the pulses with $\phi=0$ as shown in the right panel of Fig.~\ref{TurningCEP4590}. We take the longitudinal momentum $p_3 = -1.714$ for the Sauter pulse and $p_3 = -1.83$ for the Gaussian pulse. It shows that for the Gaussian pulse, the dominant turning points are located at $t_{p_1}=-45.69+i 10.52$ and $t_{p_2}=15.5+i 9.732$ whereas for the Sauter pulse these points are located at $t_{p_1}=-28.84+i 10.52$ and $t_{p_2}=30.57+i 10.52$. Here, for the Sauter pulse, the $\rm Im(t_{p_1})=\rm Im(t_{p_2})=10.52$. The other turning points are equidistant from the imaginary axis and the spectrum shows the oscillations which are caused by the interference among the neighbouring turning points. 
\begin{figure}[t]
          \begin{center}
           {\includegraphics[width = 2.4in]{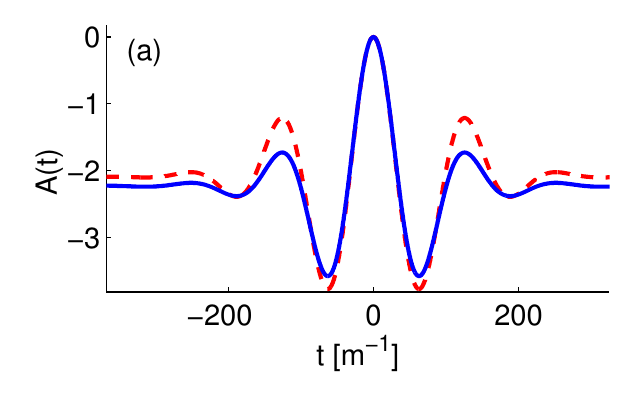}~
            \includegraphics[width = 2.4in]{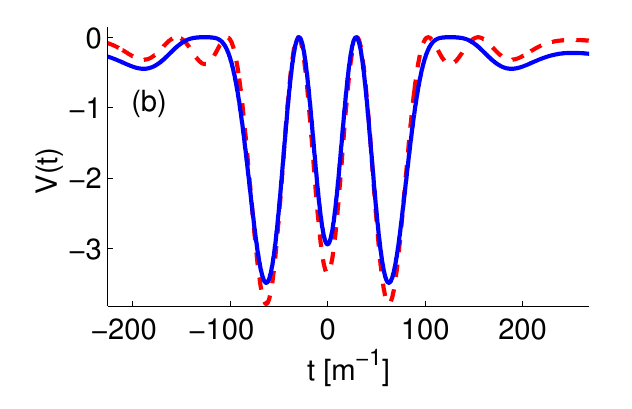}}
          \caption{Plot of $A(t)$ and $V(t)$ for Gaussian (dashed line) and Sauter (line) with $\phi = 0$. The $V(t) = -(p_3-eA(t))^2$ is plotted for the longitudinal momentum values $p_3 = -1.830$ for the Gaussian pulse and $p_3 = -1.7143$ for the Sauter pulse. All the units are taken in electrons mass unit. The field parameters are $E_0 = 0.1$, $\tau = 100$, and $\omega = 0.05$.}
          \label{VACEP90}
            \end{center}
\end{figure}   
The result is consistent with the scattering potentials for both the pulses with CEP $\phi=0$. Fig.~\ref{VACEP90} shows the vector potential $A(t)$ (Fig.~\ref{VACEP90} (a)) and the scattering potential $V(t)$ (Fig.~\ref{VACEP90} (b)). Both  $A(t)$ and $V(t)$ are symmetric about $t=0$ and we take the longitudinal momentum $p_3 = -1.830$ for the Gaussian pulse and $p_3 = -1.7143$ for the Sauter pulse. The scattering potential shows multiple bumps with comparable potential strength and hence reflection amplitude is the sum of the all the reflected wave by the scattering potential bumps. Therefore the reflectance coefficient which is related to the LMS of the created particles shows the oscillations due to the interference of the reflected waves.

\subsection{LMS for Sauter and Gaussian pulses with frequency chirping}\label{Result_Chirp}
\begin{figure}[t]
                   \begin{center}
                   {\includegraphics[width = 2.0in]{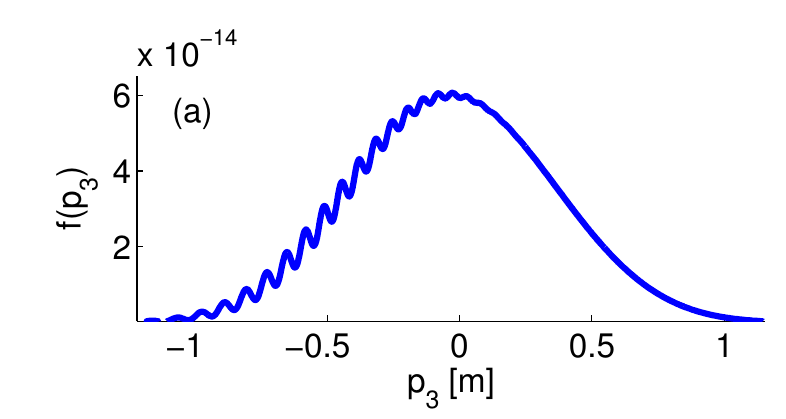}
                   \includegraphics[width = 2.0in]{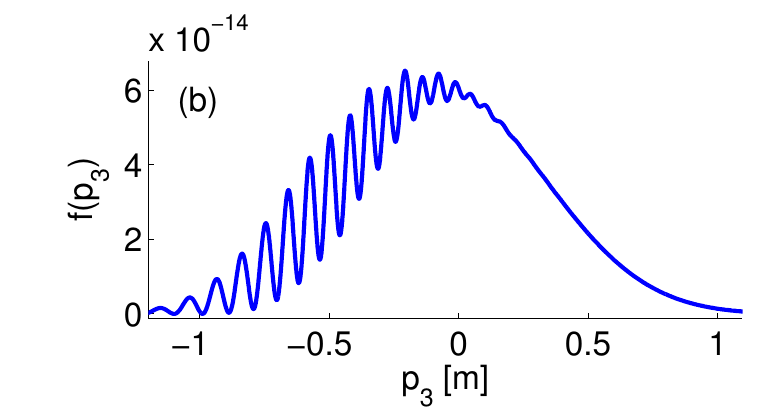}
                   \includegraphics[width = 2.0in]{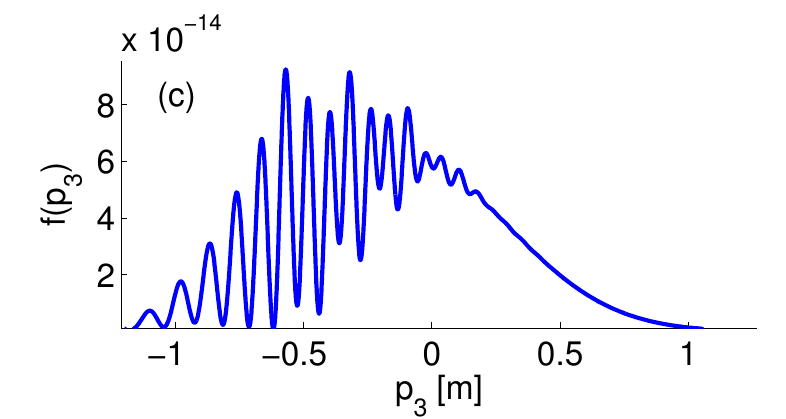}} 
                   \caption{Asymptotic distribution function of the created particles in the presence of time dependent oscillating electric field with  Sauter pulse as a function of longitudinal momentum of the particles for different values of linear frequency chirp parameter $(\beta)$. (a) $\beta=0.00025$, (b) $\beta=0.0005$, (c) $\beta=0.00075$. The transverse  momentum of the created particle is taken to be zero ($\textbf{p}_{\perp} = 0$) and all the units are taken in electronic mass unit. The field parameters are $E_0 = 0.1$, $\tau = 100$, central frequency $\omega \tau = 5$ and CEP $\phi = \pi/2$ .}
                   \label{SauMomchirp}
                   \end{center}
               \end{figure}

Frequency chirping is an important parameter for ultrashort pulses and it has resulted in many interesting effects for pair production by different types of EM fields, see for example Ref.~\cite{Chirp1DQSu} and references therein. The effect of the frequency chirping of the temporal pulse profile in the LMS of the created particles has been studied in \cite{PhysRevD.82Dumlu,Chirp1DQSu}. Here we use the expression of the electric field for the ultrashort pulses with the linear frequency chirp parameter $\beta$ which is given by $E(t) = E_0g(t)\sin(\omega t+ \beta t^2 + \phi)$ with $g(t)$ being either a Gaussian or Sauter envelope. The presence of $\beta$  modifies the frequency in a time dependent way -- for negative times the effective frequency is lower while for positive times the effective frequency is higher. As discussed in the previous subsection the tail regions of the two pulse forms may differ significantly in presence of frequency chirping and hence give rise to different LMS of created particles at asymptotic times. In order to verify this claim, we plot the  LMS as seen in Fig.~\ref{SauMomchirp} for  oscillating electric field with Sauter pulse with $\omega \tau = 5$  and the value of $\beta = 0.00025$, $0.0005$, $0.00075$. We compare our results with those obtained with the corresponding Gaussian pulses reported in Fig.~3 of Ref.~\cite{PhysRevD.82Dumlu}. For a small value of linear chirp $\beta = 0.00025$, the asymptotic LMS for the Sauter pulse shows small oscillation over the smooth profile for a negative value of $p_3$. For the Gaussian pulse, the oscillations in the spectrum are much more pronounced and the spectrum is not centred at $p_3=0$, see the top left plot of Fig.~3 in Ref.~\cite{PhysRevD.82Dumlu}. As we increase the value of $\beta$, the shape of the distribution for the Sauter pulse remains intact although the oscillation amplitude gets enhanced. However, for the Gaussian pulse, it has been shown in \cite{PhysRevD.82Dumlu} that the LMS becomes highly oscillating with the irregular profile.  

\subsection{LMS in the multi-photon regime}
\begin{figure}[h]
\begin{center}
\includegraphics[width = 4.5in]{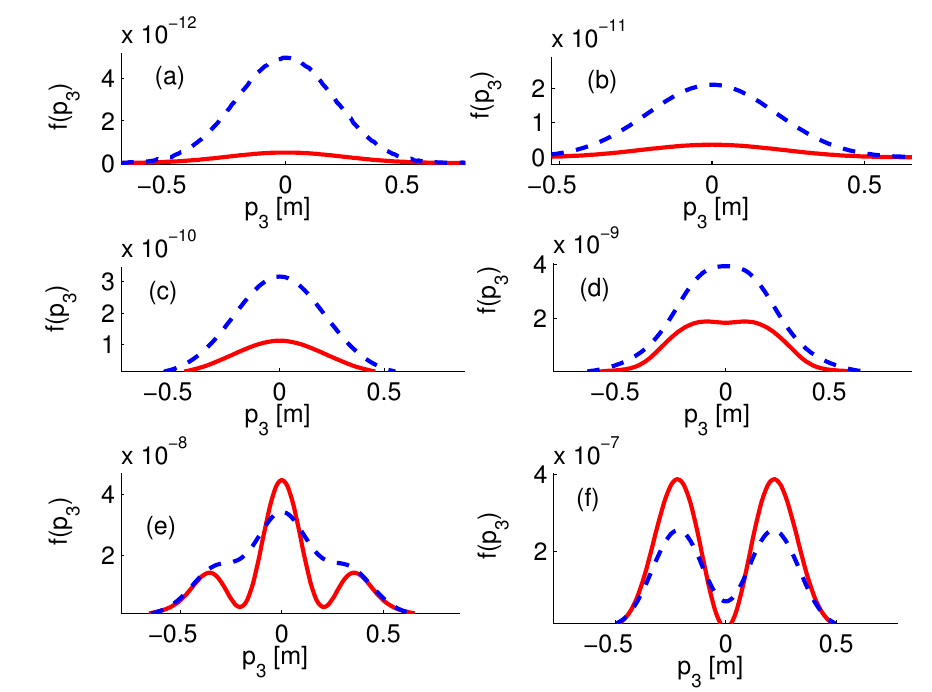}
\caption{Longitudinal LMS of created particles in the presence of time dependent smooth and oscillating electric field with Sauter (blue dashed line) and Gaussian pulses (red solid line) for (a) smooth, (b) $\omega \tau = 1$, (c) $\omega \tau = 2$, (d) $\omega \tau = 3$, (e) $\omega \tau = 4$, (f) $\omega \tau = 5$ in the intermediate and multi-photon regimes at $t=10\tau$. The value of transverse momentum is taken to be zero and all the units are taken in electrons mass unit. The field parameters are $E_0 = 0.1$, $\tau = 10$, and CEP $\phi = \pi/2$.}
   \label{MultiPhotoGauSau}
   \end{center}
   \end{figure}
LMS in the multi-photon regime for the Gaussian pulse was reported in Ref.~\cite{PhysRevAMocken,FIPTMultiphotonSmolyansky}. It was shown that spectrum takes the shapes of multi-modal function in this regime. In order to explore the effect of the temporal profile on the LMS in this regime, we analyse the spectrum of created particles in multi-photon regime for both Gaussian and Sauter pulses. We choose the parameters of laser pulse in such a way that the Keldysh parameter $\xi$ varies from $\xi\ll 1$ to $\xi\gg 1$. Fig.~\ref{MultiPhotoGauSau} shows the longitudinal LMS for both the pulses for the smooth profile (governed by the temporal envelope function as $E(t) = E_0\exp(-t^2/{2\tau^2})$ for the Gaussian pulse and $E(t) = E_0\cosh^{-2}(t/\tau)$ for the Sauter pulse) and the oscillating electric field  pulses with a few-cycle oscillations $\omega\tau =1, 2, 3, 4$ and $5$ for $t=10\tau$.
In Fig.~\ref{MultiPhotoGauSau}(a), it is shown for the smooth Sauter and Gaussian pulses with $\tau = 10$ and $E_0 = 0.1$. The Keldysh parameter in this case is close to $1$. It gives an intermediate regime of pair production via tunnelling and multi-photon processes which is known as the nonperturbative multi-photon regime. The spectrum has a unimodal profile centred at $p_3 = 0$ and its peak value is larger for the Sauter pulse than that for the Gaussian pulse.  This unimodal profile persists for the pulses having  a few-cycle oscillation upto $\omega\tau = 3$ (see Figs.~\ref{MultiPhotoGauSau}(b-d)). The interference of multi-photon processes with different photon numbers is found for the oscillating electric field  Gaussian pulse with $\omega\tau = 4$. The LMS, in this case, has a trimodal profile as seen in Fig.~\ref{MultiPhotoGauSau}(e). The central peak which is  located at $p_3 = 0$ is much more prominent than the two symmetrical side peaks which are located at $p_3 = \pm 0.37$. However for the Sauter pulse the spectrum is still unimodal except for the slight modulation. The peak value of the spectrum for the Gaussian pulse is larger than that for the Sauter pulse. Fig.~\ref{MultiPhotoGauSau}(f) shows the spectrum for $\omega\tau = 5$ for both the pulses. The profile is bimodal wherein the peaks are symmertically located about $p_3=0$ at $p_3 = \pm 0.25$. Its peak value, in this case too, is larger for the Gaussian pulse.

\section{conclusion}\label{conclusion}
To conclude, the effect of the temporal pulse shape of intense ultrashort pulses on the longitudinal momentum distribution of $e^+e^-$ pairs is studied using quantum kinetic equation. It is shown that the distribution is quite sensitive to the temporal profile -- to the extent that the two closely resembling temporal envelopes namely,  Gaussian and Sauter with the same pulse parameters give rise to significantly different LMS of pairs at all the temporal stages of evolution. The temporal stages are classified, for convenience, into two distinct regions -- transient and asymptotic regions. It is found that the transient region for the Gaussian pulse has a larger temporal extent.  In the transient region, the spectrum is smooth with a single peak for both the pulses. However, the location of the peak, the peak height and the width are different for any instant of time and they evolve differently with time. However, the peak height for the Gaussian pulse is consistently higher than that for the Sauter pulse. 

In the other regime, where the spectrum does not change with time, the peak position of the spectrum nearly coincides in both the cases. However, in the asymptotic region, just contrary to the trend in the transient region, the peak height of the spectrum for the Gaussian pulse is lower than that for the Sauter pulse as long as the number of a few-cycle oscillations, $\omega \tau < 5$. The LMS for the oscillating electric field with Gaussian shows oscillations over the smooth profile due to the quantum mechanical interference of the multiple reflections of quasi-particle waves from bumpy time-dependent potential. These oscillations are suppressed in the case of Sauter pulse. The onset of oscillation takes place at $\omega\tau = 6$ for the Sauter pulse compared to $\omega\tau = 4$ for the Gaussian pulse. Furthermore, for the same value of $\omega \tau$ the amplitude of oscillation is smaller for the Sauter pulse. In fact, it is due to this interference effect that the peak height of the LMS for the Gaussian pulse takes over that for the Sauter pulse for $\omega \tau \geq 5$. The sensitivity of the LMS to the temporal pulse forms is explained by analyzing the shape of potential causing over barrier reflections and also the turning point structure in complex time plane for the two pulse forms.

The differences in the asymptotic time LMS of the two pulses get much more prominent on increasing the linear frequency chirp in these pulses and also on varying the CEP. Furthermore, the profile and the location of the spectrum is vastly different for the two pulses.

The foregoing observation about the sensitivity of the LMS on the temporal pulse profile not only holds for the tunneling regime of the pair production, but it also persists in the intermediate and the multiphoton regimes.

 After the manuscript was submitted we came across the recent publication by Schutzhold et. al. \cite{Torgrimsson2017pzs} where the sensitivity of temporal pulse profile is shown for pair production by dynamically assisted Schwinger mechanism. It has been pointed out that the Fourier transforms in the complex (frequency) plane of envelopes is fairly different, even while the envelopes look similar on the real (time) axis. As pointed out earlier, it may be possible in the foreseeable future to generate such intense electric fields at the focal region of the counter-propagating laser beams. However, away from the focus the spatial dependence of the laser electromagnetic field no longer be neglected and one has to use more sophisticated methods, like DWH formula, WKB and instanton method  \cite{RalfScheutzholdDSchw,Kohlfurst1PhysRevD,Kohlfurst2PhysRevD,ORTHABER201180,RalfSchneider2014mla,OttoKampfer2014ssa,Scheutzhold2012,ScheutzholdPhysRevLett,CatalysisGiesPhysRevD,OttoKampferPhysRevD,ScheutzholdPulseShapePhysRevD,Panferov2015yda,ScheutzholdPrefactorPhysRevD,Torgrimsson2017pzs}. 
 
Although it may appear somewhat far fetched, measuring LMS of the pairs may provide a possible method for the determination of the temporal profile for ultrashort pulses.

\section*{ACKNOWLEDGMENTS}
 CB gratefully acknowledge the financial support from Homi Bhabha National Institute (HBNI) for carrying out this research work. Authors would like to thank Prof. Arup Banerjee and Prof. Aparna Chakrabarti for helpful discussions and encouragement during this work.

 \appendix
\section{Onset of oscillations in the LMS of oscillating electric field with Sauter and Gaussian pulse profiles}\label{Appendix}
In this appendix, we use Eq.~\ref{eq:N_k(inf)} to determine the onset of oscillations for oscillating electric field with Sauter and Gaussian pulse profiles. We first evaluate  the integrals $K^{(p)}_{\textbf{p}} = \bigg|\int\limits_{t^*_p}^{t_p}dt~\omega(\textbf{p}, t)\bigg|$, $\Theta^{(p,p^{\prime})}_\textbf{p} = \int\limits_{Re(t_p)}^{Re(t^{\prime}_p)}dt ~\omega(\textbf{p}, t)$  and hence $f_{\textbf{p}}(\infty)$ for $\omega\tau = 4$ and $6$  for both pulses. 
We take three pairs of turning points (the central one and the adjacent ones on either side of the central pair), $t_{p1}$, $t_{p2}$, and $t_{p3}$ and their complex conjugates. So we have from Eq.~\ref{eq:N_k(inf)} for the three pairs of turning points
\begin{equation}\label{eq:N_asymp_3TP}
\begin{split}
f_{\textbf{p}}(\infty) \approx e^{-2K^{(p1)}_{\textbf{p}}}+e^{-2K^{(p2)}_{\textbf{p}}}+e^{-2K^{(p3)}_{\textbf{p}}}-2\cos\bigg(2\Theta^{(p1,p2)}_\textbf{p}\bigg)~e^{-K^{(p1)}_{\textbf{p}}-K^{(p2)}_{\textbf{p}}}-2\cos\bigg(2\Theta^{(p2,p3)}_\textbf{p}\bigg)~e^{-K^{(p2)}_{\textbf{p}}-K^{(p3)}_{\textbf{p}}}\\+2\cos\bigg(2\Theta^{(p1,p3)}_\textbf{p}\bigg)~e^{-K^{(p1)}_{\textbf{p}}-K^{(p3)}_{\textbf{p}}},
\end{split}
\end{equation}
where the three pairs of turning points are taken from left to right. In Fig.~\ref{GauSauterTurning1May2018} the values of turning points for $\omega\tau = 4$ are $t_{p1} = -83.1+13.7068 i$, $t_{p2} = 0+9.7359 i$, $t_{p3} = 83.1+13.7068 i$ for the Gaussian pulse,  whereas for the Sauter pulse the values of turning points are  $t_{p1} = -87.056+19.5236 i$, $t_{p2} = 0+9.721 i$, $t_{p3} = 87.056+19.5236i $. The values of the integrals for the Sauter pulse are $K^{(p1)}_{\textbf{p}} = \bigg|\int\limits_{t^*_{p1}}^{t_{p1}}dt~\omega(\textbf{p}, t)\bigg| = 32.3513$, $K^{(p2)}_{\textbf{p}} = \bigg|\int\limits_{t^*_{p2}}^{t_{p2}}dt~\omega(\textbf{p}, t)\bigg| = 15.3754$, and $K^{(p3)}_{\textbf{p}} = \bigg|\int\limits_{t^*_{p3}}^{t_{p3}}dt~\omega(\textbf{p}, t)\bigg| = 32.3513$. Therefore, in  Eq.~\ref{eq:N_asymp_3TP} the value of the exponentials are  $e^{-2K^{(p1)}_{\textbf{p}}} = 7.9438\times 10^{-29}$, $e^{-2K^{(p2)}_{\textbf{p}}} = 4.4165\times 10^{-14}$, $e^{-2K^{(p3)}_{\textbf{p}}} = 7.9438\times 10^{-29}$, $e^{-K^{(p1)}_{\textbf{p}}-K^{(p2)}_{\textbf{p}}} = 1.87309\times 10^{-21}$, $e^{-K^{(p2)}_{\textbf{p}}-K^{(p3)}_{\textbf{p}}} = 1.87309\times 10^{-21}$, and $e^{-K^{(p1)}_{\textbf{p}}-K^{(p3)}_{\textbf{p}}} = 7.9438\times 10^{-29}$. The orders of the exponentials show that the main contribution to $f_{\textbf{p}}$ comes from only one turning point pair, $t_{p2}$ and its conjugate, which lie at the centre. The LMS, therefore, is unimodal with  no interference effect due to reflections from other turning points, for $\omega\tau = 4$ for the Sauter pulse. Similar calculation for the Gaussian pulse gives:  
 $K^{(p1)}_{\textbf{p}} = 21.8979$, $K^{(p2)}_{\textbf{p}} = 15.3933$, and $K^{(p3)}_{\textbf{p}} = 21.8979$ and the value of the exponentials are  $e^{-2K^{(p1)}_{\textbf{p}}} = 9.5448\times 10^{-20}$, $e^{-2K^{(p2)}_{\textbf{p}}} = 4.26148\times 10^{-14}$, $e^{-2K^{(p3)}_{\textbf{p}}} = 9.5448\times 10^{-20}$, $e^{-K^{(p1)}_{\textbf{p}}-K^{(p2)}_{\textbf{p}}} = 6.37731\times 10^{-17}$, $e^{-K^{(p2)}_{\textbf{p}}-K^{(p3)}_{\textbf{p}}} = 6.37731\times 10^{-17}$, and $e^{-K^{(p1)}_{\textbf{p}}-K^{(p3)}_{\textbf{p}}} = 9.5448\times 10^{-20}$.  In this case the main contribution to the reflection coefficient is from the central turning point. Hence the shape of the LMS is unimodal. However, the interference between the central and the adjacent turning points also appears with relative strength of about $1.5 \times 10^{-3}$.  So for the Gaussian pulse case the onset of oscillation takes place for the first time for $\omega\tau = 4$ and the LMS showing small amplitude oscillations over the unimodal profile. 
 
 Now we calculate $f_{\textbf{p}}(\infty)$ for $\omega\tau = 6$ for the Sauter pulse which shows for the first time the onset of oscillation in the LMS as seen in the left panel of Fig.~\ref{SauterMomOmTau}. Here the values of the turning points for $p_3 = 0$ are $t_{p1} = -54.31278+12.2396 i$, $t_{p2} = 0+9.453768 i$, $t_{p3} = 54.31278+12.2396 i$ and their complex conjugates. The values of the integrals are $K^{(p1)}_{\textbf{p}} = 19.7175$, $K^{(p2)}_{\textbf{p}} = 15.0493$, and $K^{(p3)}_{\textbf{p}} = 19.7175$. The value of the exponential are $e^{-2K^{(p1)}_{\textbf{p}}} = 7.47409\times 10^{-18}$, $e^{-2K^{(p2)}_{\textbf{p}}} = 8.47864\times 10^{-14}$, $e^{-2K^{(p3)}_{\textbf{p}}} = 7.47409\times 10^{-18}$, $e^{-K^{(p1)}_{\textbf{p}}-K^{(p2)}_{\textbf{p}}} = 7.96105\times 10^{-16}$, $e^{-K^{(p2)}_{\textbf{p}}-K^{(p3)}_{\textbf{p}}} = 7.96105\times 10^{-16}$, and $e^{-K^{(p1)}_{\textbf{p}}-K^{(p3)}_{\textbf{p}}} = 7.47409\times 10^{-18}$. So the value of the phase integrals for the dominant terms are $\Theta^{(p1,p2)}_\textbf{p} = \int\limits_{Re(t_{p1})}^{Re(t_{p2})}dt ~\omega(\textbf{p}, t) = -81.5475$, $\Theta^{(p2,p1)}_\textbf{p} = \int\limits_{Re(t_{p2})}^{Re(t_{p3})}dt ~\omega(\textbf{p}, t) = 81.5475$ and $\cos\bigg(2\Theta^{(p1,p2)}_\textbf{p}\bigg) = \cos\bigg(2\Theta^{(p2,p3)}_\textbf{p}\bigg) = 0.9643$. Therefore the interference term $2\cos\bigg(2\Theta^{(p1,p2)}_\textbf{p}\bigg)e^{-K^{(p1)}_{\textbf{p}}-K^{(p2)}_{\textbf{p}}}+2\cos\bigg(2\Theta^{(p2,p3)}_\textbf{p}\bigg)e^{-K^{(p2)}_{\textbf{p}}-K^{(p3)}_{\textbf{p}}} = 3.0708\times 10^{-15}$ becomes comparable to  the term $e^{-2K^{(p2)}_{\textbf{p}}} = 8.47864\times 10^{-14}$, representing the reflection from the central turning point. Here the modulation over the unimodal profile appears with the relative strength of $3.6 \times 10^{-2}$. The peak value of the distribution $f_{\textbf{p} = 0}(\infty) \approx 8.17154\times 10^{-14}$.


\end{document}